\documentclass{aa}  
\usepackage[varg]{txfonts}
\usepackage{graphicx}
\usepackage{siunitx}
\usepackage{xcolor}
\usepackage{natbib}
\usepackage{booktabs}
\usepackage{boldline}
\usepackage{nccmath}
\usepackage{amsmath}
\usepackage{tabularx}
\usepackage{algorithm2e}
\usepackage{csquotes}
\usepackage{comment}
\usepackage{gensymb}
\usepackage{enumitem}
\usepackage{float}
\usepackage{subcaption}
\usepackage[labelfont=bf]{caption}
\usepackage{mwe}
\usepackage[colorlinks,citecolor=blue,urlcolor=blue,bookmarks=false,allcolors=blue,hypertexnames=true]{hyperref} 
\usepackage{threeparttable}
\usepackage{color}
\usepackage[colorinlistoftodos]{todonotes}
\usepackage{tabularx}
\usepackage{adjustbox}
\usepackage{blindtext, rotating}
\usepackage{array,booktabs}

\newcommand\clearrow{\global\let\rowmac\relax}
\makeatletter

\DeclareRobustCommand{\OI}{%
  \mbox[{O\check@mathfonts\fontsize\sf@size\z@\selectfont I}]%
}
\DeclareRobustCommand{\OII}{%
  \mbox[{O\check@mathfonts\fontsize\sf@size\z@\selectfont II}]%
}
\DeclareRobustCommand{\OIII}{%
  \mbox[{O\check@mathfonts\fontsize\sf@size\z@\selectfont III}]%
}
\makeatother
\def \kms         {km$\,$s$^\mathrm{-1}$}

\def \arcsec      {\text{$^\mathrm{\prime\prime}$}}

\def \whz         {\,W\,Hz$^{-1}$}

\def \ergs    {$\mathrm{\times 10^{-17} erg s^{-1} cm^{-2} \AA^{-1}}$}
\def \ergsunit    {$\mathrm{erg s^{-1} cm^{-2} \AA^{-1}}$}
\defcitealias{Gereb2014}{G14}
\defcitealias{Gereb2015}{G15}
\defcitealias{Maccagni2017a}{M17}

\newcommand{\hbeta}{\text{H\textsc{$\beta$}}}
\newcommand{\halpha}{\text{H\textsc{$\alpha$}}}
\newcommand{\hgamma}{\text{H\textsc{$\gamma$}}}
\newcommand{\hdelta}{\text{H\textsc{$\delta$}}}

\newcommand{\Nii}{\text{[N\textsc{ii}]}}
\newcommand{\Sii}{\text{[S\textsc{ii}]}}

\newlist{inlineroman}{enumerate*}{1}
\setlist[inlineroman]{itemjoin*={{ }},afterlabel=~,label=\roman*)}

\makeatletter
\renewcommand{\fnum@figure}{Figure \thefigure}
\makeatother

\begin{document} 
\title{Ionised gas outflows over the radio AGN life cycle}
\author{Pranav Kukreti\inst{1,2}\thanks{\email{kukreti@astro.rug.nl}},
Raffaella Morganti\inst{2,1}, Clive Tadhunter\inst{3} and Francesco Santoro\inst{2}
}

\authorrunning{Kukreti et al.}
\titlerunning{Ionised gas outflows over the radio AGN life cycle}

\institute{
Kapteyn Astronomical Institute, University of Groningen, Postbus 800, 9700 AV Groningen, The Netherlands
\and
ASTRON, the Netherlands Institute for Radio Astronomy, Oude Hoogeveensedijk 4, 7991 PD Dwingeloo, The Netherlands
\and
Department of Physics and Astronomy, University of Sheffield, Sheffield S3 7RH, UK
}

  \abstract{Feedback from active galactic nuclei (AGN) is known to affect the host galaxy's evolution. In radio AGN, one manifestation of feedback is seen in gas outflows. However, it is still not well understood whether the effect of feedback evolves with the radio AGN life cycle. In this study, we aim to investigate this link using the radio spectral shape as a proxy for the evolutionary stage of the AGN. We used \OIII~emission line spectra to trace the presence of outflows on the ionised gas. Using a sample of uniformly selected 129 radio AGN with $L_\textrm{1.4\,GHz}\approx10^{23}-10^{26}$\whz, and a mean stacking analysis of the \OIII~profile, we conclude that the ionised gas outflow is linked to the radio spectral shape, and it evolves with the evolution of the radio source. We find that sources with a peak in their radio spectra (optically thick), on average, drive a broad outflow ($FWHM\approx1330\pm418$\,\kms) with a velocity $v_\textrm{out}$\,$\approx$\,240\,\kms. However, we detect no outflow in the stacked \OIII~profile of sources without a peak in their radio spectrum (optically thin). We estimate a mass outflow rate of $0.09-0.41$\,M$_{\odot}$\,yr$^{-1}$, and a kinetic power of $0.1-1.8\times10^{41}$\,erg\,s$^{-1}$ for the outflow. In addition, we find that individual outflow detections are kinematically more extreme in peaked than non-peaked sources. We conclude that radio jets are most effective at driving gas outflows when young, and the outflow is typically short lived. Our stacking analysis shows no significant dependence of the presence of ionised gas outflows on the radio morphology, 1.4\,GHz luminosity, optical luminosity and Eddington ratio of these sources. This suggests that in our sample, these properties do not play a defining role in driving the impact of the nuclear activity on the surrounding gas. We also identify candidate restarted AGN in our sample, whose \OIII~profiles suggest that they have more disturbed gas kinematics than their evolved counterparts, although the evidence for this is tentative. Our findings support the picture where the impact of AGN feedback changes as the source evolves, and young radio jets interact with the ambient medium, clearing a channel of gas as they expand.}

    \date{Received 15 December 2022 / Accepted 3 May 2023}    

    \maketitle
\section{Introduction}
\label{introduction}

Active galactic nuclei (AGN) release large amounts of energy, which impacts the evolution of the host galaxy. This phenomenon, known as AGN feedback, is important for understanding galaxy evolution (for example, \citealt{Silk1998, Fabian2012,Harrison2014,McNamara2012}). Tracing the impact of the various forms of nuclear activity on the host galaxy requires observing its effect on the surrounding gas. One signature of this impact is the presence of kinematically disturbed gas and gas outflows. \par

Evidence for outflows has been found for many AGN in ionised, neutral atomic, and molecular gas (for example, \citealt{Cicone2014,Garcia-Burillo2014,Harrison2018,Husemann2019,Maksym2022, Morganti2018,Muller-Sanchez2011,Falcao2021,Veilleux2013,Veilleux2020,Winkel2022}). Broadly, two mechanisms have been suggested to explain the origin of the outflows. The first mechanism is radiative or quasar mode feedback, which occurs when gas is ejected in winds due to radiation pressure from the nucleus. The second mechanism is mechanical or radio mode feedback, which occurs due to radio jets' interaction with the surrounding gas (for example, \citealt{Fabian2012,King2015}). Both mechanisms are likely to play a role in driving feedback in different AGN. Indeed, radiative mode is expected to be dominant in high luminosity sources, close to Eddington, whereas mechanical mode is found more often in sources with luminosity much lower than Eddington. Therefore, it is important to quantify their impact in different AGN. Another important parameter is the changes in the impact of feedback during the AGN lifetime. This would provide a better understanding of AGN feedback, and it is also an important input for numerical simulations of galaxy evolution.\par
\begin{figure*}
  \includegraphics[width=2\columnwidth]{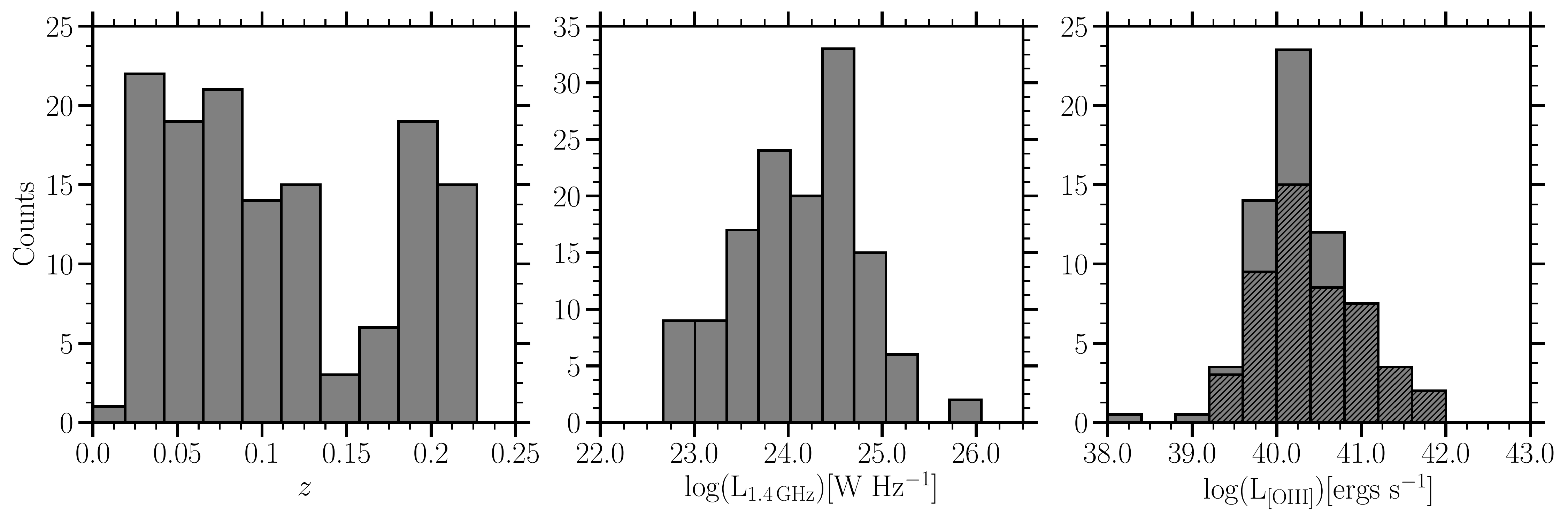}
  \caption{Sample properties. (\textit{left}) Redshift and (\textit{middle}) peak 1.4\,GHz radio luminosity distributions of the 135 sources in the sample (including mergers). The dip in the number of sources in the interval $0.136<z<0.175$ is due to the exclusion of sources affected by RFI \citepalias{Gereb2014}. (\textit{right}) \OIII~luminosity distribution for all the sources. For non-detections, we used the upper limit of the \OIII~luminosity. The hatched area shows the distribution for only the \OIII~detections.}
  \label{sample_properties_distribution}    
\end{figure*}

In radio AGN, many studies have found evidence for jet-driven feedback in AGN host galaxies (for example, \citealt{Husemann2019,Jarvis2019,Morganti2021a,Murthy2022,Nesvadba2007,Nesvadba2021,Ruffa2022,Schulz2021,Venturi2021} for some recent results). Some trends have also been found for the possible impact of jets on the surrounding gas. For example, in a sample of optically selected AGN, \citet{Mullaney2013} 
found that 1.4\,GHz radio luminosity had the most influence on the ionised gas kinematics of the narrow line region, compared to the bolometric luminosity. Therefore, they could disentangle the impact of jets from radiation. A positive link between young radio AGN (and thus a young jet) and disturbed gas kinematics has also been suggested by both ionised and neutral gas studies (\citealt{Holt2008,Gereb2015,Roche2016,Maccagni2017a,Molyneux2019,Murthy2019,Santoro2020}), which is in line with the picture where feedback evolves over the AGN lifetime. This is supported by numerical simulations of an interaction between jets and a clumpy medium, which allowed authors of previous studies to conclude that the impact changes as the jets evolve to larger scales \citep{Sutherland2007,Mukherjee2018,Meenakshi2022a}. \par
While \citet{Mullaney2013} could disentangle the effect of radiative and mechanical mode feedback, emphasising the importance of the latter for driving gas outflows, the connection to young radio AGN is still mostly limited to powerful sources or small groups. Furthermore, most of the studies use high frequency (GHz range) radio morphology as a proxy for the evolutionary stage of radio AGN (for example, \citealt{Holt2008,Mullaney2013,Molyneux2019}). However, as we explain below, radio spectral properties can give more accurate information about the evolutionary stage of the AGN. \par

In this paper, we investigate a sample of radio sources for which we characterise the evolutionary stage using the radio spectral information, extended to low frequency ($\sim$150\,MHz) by LOFAR. We combine this with the properties of ionised gas provided by SDSS, thus studying the presence of outflows and quantifying the feedback effect. The sample size allows us to perform a stacking analysis to investigate the presence of outflows in groups of sources selected based on their optical and radio properties, so to disentangle the relevance and effect of various key parameters (for example age of the radio source, Eddington rate, excitation state etc). \par

Over the years, radio AGN have been observed in phases of activity and dormancy, which is collectively called the radio AGN life cycle (see \citealt{Morganti2017b} for a review). Young radio AGN are expected to show a peak in their radio spectrum (usually in the GHz frequency range), due to absorption of the synchrotron emission, and bright jets on small scales (sub-kpc to kpc). The absorption is either due to synchrotron self-absorption (SSA) or free–free absorption (FFA). This phase lasts $10^{5}-10^{6}$\,Myr and has been detected in peaked spectrum sources (and in particular compact steep spectrum, CSS, and gigahertz peaked spectrum, GPS, sources; \citealt{Slob2022ExtragalacticGalaxies} and see \citealt{ODea1998} and \citealt{ODea2021a} for a review). As the source evolves and the jets grow, the peak in the radio spectrum shifts to lower frequencies, eventually below the observing frequencies of most radio telescopes. At this stage, the radio spectra is observed to be optically thin. However, with low frequency radio telescopes such as the LOFAR telescope, it is possible to detect the peak down to $\sim$\,150\,MHz, allowing us to detect peaked sources that would be missed in higher frequency data. \par

After a phase of activity which can last up to $10^{8}$\,yr, the fuelling of the radio jets stops and there is no more injection of fresh electrons. This can lead to a remnant phase with low surface brightness diffuse emission, and a steep radio spectra, due to preferential radiative cooling at high frequencies ($\sim$GHz, for example, \citealt{Saripalli2012b,Brienza2017,Jurlin2021}). Since the lower frequency end of the radio spectrum is affected last, LOFAR is an ideal instrument to detect extended radio emission from AGN, even in the remnant phase. Such low frequency data is also useful to accurately characterise source morphology. In some cases (13-15\%), a new phase of activity can be detected even before the radio plasma from the older phase has faded below detection limits \citep{Jurlin2020}. These sources are called restarted radio AGN (for example, \citealt{Parma2007a,Shulevski2015a,Brienza2018,Morganti2021c,Morganti2021b}, and see \citealt{Saikia2009} for a review). These sources are expected to have a young phase of activity in the centre, characterised by a peaked spectrum, surrounded by diffuse emission from an older phase of activity (for example in 3C\,293 by \citealt{Kukreti2022a}). Therefore, the radio spectral shape can be used as a good tracer for the AGN life cycle.\par

Tracing the radio spectral shape accurately requires a wide frequency coverage. To achieve this, we used a combination of radio surveys at three frequencies - the LOFAR Two-metre Sky Survey (LoTSS; \citealt{Shimwell2017}) at 144\,MHz, the Faint Images of the Radio Sky at Twenty-cm survey (FIRST; \citealt{Becker1995TheCentimeters}) at 1400\,MHz and the Very Large Array Sky Survey (VLASS; \citealt{Lacy2016}) at 3000\,MHz. These surveys were also chosen for their high angular resolution (6\arcsec for LoTSS, 5.4\arcsec for FIRST, and $\sim$2.5\arcsec for VLASS), which ensures that we trace the spectral shape of the central few kpc ($\approx$\,6-12\,kpc at the mean redshift of the sample), where the emission is expected to be dominated by the radio jets. This is needed in order to relate the radio spectral shape to the evolutionary stage of the radio jet. The resolution of these surveys is also comparable to SDSS fibre size ($\sim$3\arcsec), which allows us to compare the radio and optical emission over similar spatial scales\par
\begin{table*}
\centering
\begin{threeparttable}
\caption[]{Sample properties}
         \label{sample_properties}
\renewcommand{\arraystretch}{1}         
\setlength{\tabcolsep}{4pt}
\begin{tabular}{ccccccc}
            \hlineB{4.5}
            
    \noalign{\smallskip}
     & No. of sources  & $z$ & log($L_\mathrm{1.4\,GHz}$) & log($L_\mathrm{[OIII]}$) &  No. of \OIII~detections\\
    \noalign{\smallskip}
    \hlineB{4.5}
    
    \noalign{\smallskip}
    
    All radio AGN & 129 & 0.12 & 24.6 & 40.9 & 93 \\
    
    \hline
    \noalign{\smallskip}
    \textit{Radio spectral shape} \\
    Peaked & 37 & 0.10 & 24.4 & 40.9 & 30 \\
    Non-peaked & 44 & 0.12 & 24.9 & 41.2 & 32 \\    
    Flat-spectrum & 43 & 0.12 & 24.4 & 40.5 & 27 \\   
    Convex & 5 & 0.16 & 24.6 & 41.0 & 4 \\ 
    \noalign{\smallskip}
    \hline
    \noalign{\smallskip}
    \textit{Radio morphology} \\
    Compact & 64 & 0.11 & 24.67 & 41.1 & 52 \\
    Extended & 65 & 0.12 & 24.61 & 40.7 & 41 \\
    \noalign{\smallskip}
    \hline
    \noalign{\smallskip}
    \textit{Central radio luminosity} \\
    $22.7 \leq \textrm{log}(L_\mathrm{1.4\,GHz}) < 24.2$ & 64 & 0.07 & 23.8 & 40.9 & 56 \\     
    $24.2 \leq \textrm{log}(L_\mathrm{1.4\,GHz})\leq 26.1$ & 65 & 0.16 & 24.9 & 41.1 & 37 \\
    \noalign{\smallskip}
    \hline
    \noalign{\smallskip}
    \textit{BPT diagram} \\
    Low-ionisation & 58 & 0.08 & 24.4 & 40.4 & 58 \\
    High-ionisation & 23 & 0.11 & 25.0 & 41.4 & 23 \\    
    \noalign{\smallskip}
    \hline
    \noalign{\smallskip}
    \textit{Optical luminosity} \\
    $38.3 \leq \textrm{log}(L_\mathrm{[OIII]})\leq 40.2$ & 65 & 0.09 & 24.3 & 39.8 & 43\\  
    $40.2 < \textrm{log}(L_\mathrm{[OIII]})\leq 42.4$ & 64 & 0.14 & 24.8 & 41.1 & 50\\
    \noalign{\smallskip}
    \hline
    \noalign{\smallskip}
    \textit{Eddington ratio} \\
    $4.6\times10^{-5} \leq \lambdaup_\textrm{Edd}\leq 1.1\times10^{-3}$ & 61 & 0.11 & 24.4 & 40.1 & 44\\ 
    $1.1\times10^{-3} < \lambdaup_\textrm{Edd}\leq 2.6\times10^{-1}$ & 60 & 0.13 & 24.8 & 40.8 & 43\\
    \noalign{\smallskip}
 
            \hlineB{4.5}
 \end{tabular}      
    \small{\textbf{Note.} Properties of the radio AGN sample and the classification groups. $z$, log($L_\mathrm{1.4\,GHz}$) and log($L_\mathrm{[OIII]}$) columns list the average SDSS redshift, average central 1.4\,GHz luminosity in W\,Hz$^{-1}$ (estimated using the FIRST peak flux density) and average observed \OIII~luminosity in erg\,s$^{-1}$ respectively. See Section~\ref{methods} for details of how the groups were constructed.}
\end{threeparttable} 
\end{table*}

Our goal is to connect the AGN life cycle to feedback. To trace the feedback on the ionised gas and its kinematic properties, we used the \OIII$\mathrm{\lambdaup\lambdaup}$4958,5007\AA~doublet emission line. We used the \OIII~line because it is sensitive to the impact of radiation and jets and therefore a good tracer for outflows. It is a forbidden transition line, and therefore can only be produced in the low density environments of the narrow line region (NLR) clouds that surround the AGN. The NLR extends up to kpc sizes, and is thus ideal for studying AGN feedback on larger scales than the high density sub-parsec-sized broad line region. \OIII~is also a strong emission line, making it easier to detect for a large number of sources.\par
The paper is organised in the following manner: in Section~\ref{sample} we describe the sample used for the study, in Section~\ref{methods} we describe the methods to obtain the radio and optical properties, and our analysis, in Section~\ref{results} we present our results, and in Section~\ref{discussion} we discuss the ionised gas kinematics as a function of different source properties. Finally, we summarise our results in Section~\ref{summary}. In Appendix A, we provide LOFAR 144\,MHz continuum maps for sources where we detect extended emission for the first time.  Throughout the paper, we have used the $\Lambda$CDM cosmological model, with H$_\mathrm{0}$ = 70 km s$^\mathrm{-1}$ Mpc$^\mathrm{-1}$, $\Omega_\mathrm{M}=0.3$ and $\Omega_\mathrm{vac}=0.7$. At the average redshift of our sample ($z=0.11$), 1\arcsec corresponds to 2\,kpc.\par

\section{Sample description}
\label{sample}

For our study we used the sample from \citet{Maccagni2017a} (hereafter M17) and \citet{Gereb2014} (hereafter G14). The sample was constructed by cross-correlating the Sloan Digital Sky Survey catalogue(SDSS DR7; \citealt{York2000TheSummary}) with the FIRST catalogue. The sample was radio flux selected to have a peak flux density $S_\mathrm{1.4GHz}$ > 30\,mJy in the redshift range $0.02 < z < 0.23$ and it had 248 sources. This radio flux density is not the total flux of the source, but the flux of the core coincident with the SDSS counterpart.
\par 

We cross-matched this sample with the LoTSS DR2 \citep{Shimwell2022} and the VLASS \citep{Gordon2021} catalogue. The cross-match was done using Topcat's nearest neighbour routine \citet{Taylor2005} with a radius of 7\arcsec, comparable to the LoTSS resolution. After the cross-match, our final sample has 135 sources covering a radio luminosity range of $\mathrm{L}_\mathrm{1.4GHz} =10^{22.6}$-10$^{26.3}$\,W\,Hz$^{-1}$. The sample contains six sources that are known galaxy mergers, with signatures of an ongoing interaction. Since we want to investigate the gas kinematics of radio AGN only, we exclude these six sources from our sample. Finally, we have a sample of 129 radio AGN. The redshift, 1.4\,GHz radio luminosity and observed \OIII~luminosity distributions (Section~\ref{data reduction}) of the sample are shown in Fig.~\ref{sample_properties_distribution}.\par

\section{Methods}
\label{methods}

\subsection{Radio properties}
\subsubsection{Radio spectral indices}
\label{spectral indices}
The combination of the high resolution LoTSS, FIRST and VLASS surveys, gives us a unique opportunity to study the radio properties of the same central region for which we have the optical properties from SDSS, thus allowing a direct comparison between the two. To do this, we have used the peak flux densities from the surveys at the central region of the sources, to estimate the spectral indices. Throughout the paper, spectral index refers to this central region spectral index.

\par
The angular resolution of VLASS ($\sim$2.5\arcsec) is significantly higher than LoTSS (6\arcsec) and FIRST (5.4\arcsec). Therefore, we smoothed all the VLASS images for our sources using a 2D Gaussian with the task \texttt{IMSMOOTH} in CASA, to achieve an angular resolution of 5.4\arcsec,  that is  the same as FIRST. This causes a significant change in spectral index for extended sources, since they have emission that is missed if we only use the peak flux density seen by the $\sim$2.5\arcsec~VLASS beam. Finally, \citet{Gordon2021} have also reported that the VLASS flux densities are underestimated by a factor of 0.87. Therefore, we have corrected the VLASS flux densities by this factor.  We have then estimated the spectral indices from 144\,MHz to 3000\,MHz, using the convention $S\propto \nu^{\alpha}$. We estimate the spectral index errors using a quadrature combination of RMS noise and flux scale errors of the surveys. The flux scale errors used are 10\% for LoTSS, 5\% for FIRST and 10\% for VLASS. The typical error on the spectral indices are 0.05 for $\alpha^{144}_{1400}$ and 0.15 for $\alpha^{1400}_{3000}$. 

\subsubsection{Radio spectral shape classification}
\label{spectral classification}
\begin{figure}
\includegraphics[width=\columnwidth]{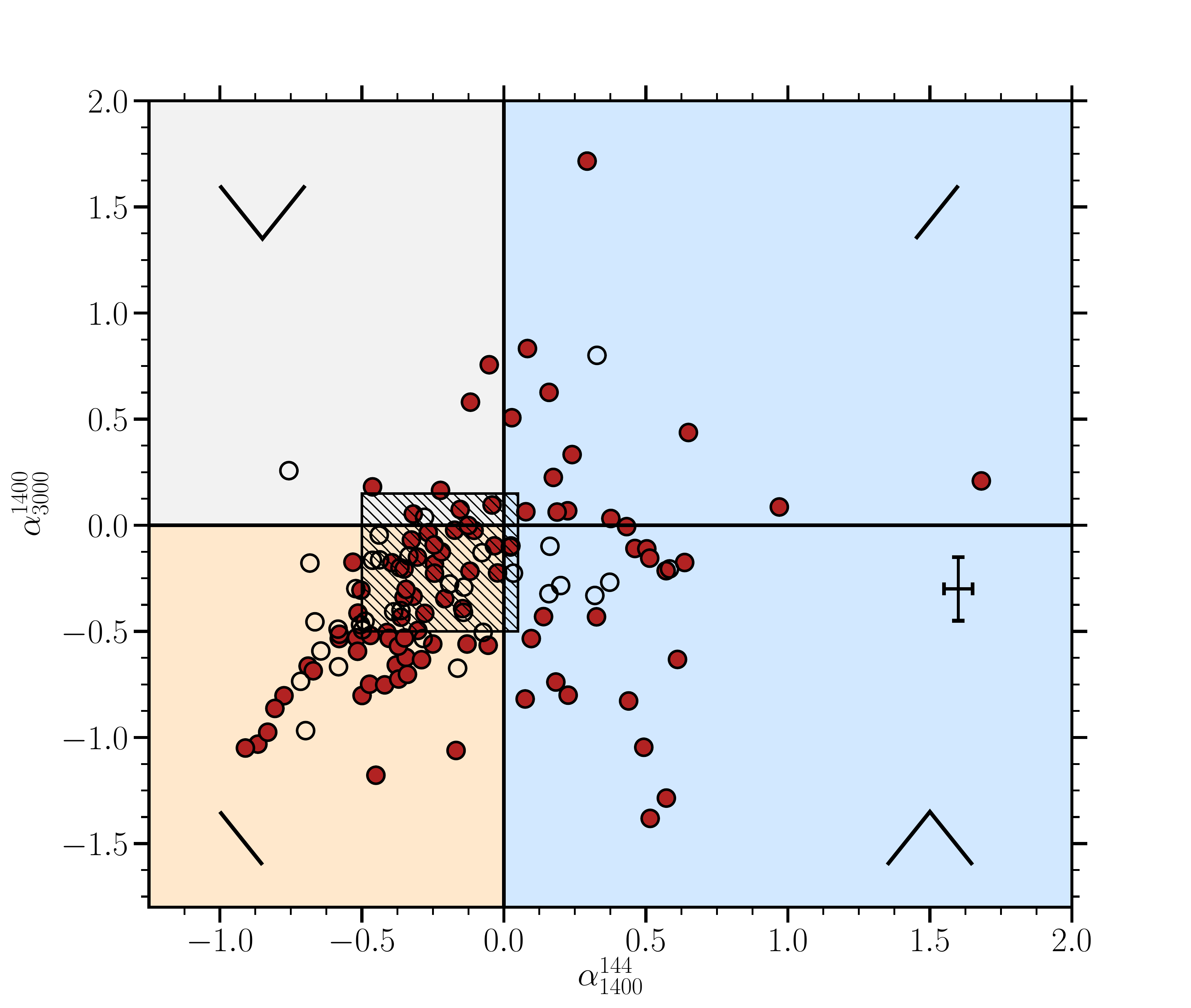}
\caption{Colour-colour plot of spectral indices for the central region of the 129 radio AGN. X-axis shows $\alpha^{144}_{1400}$ derived using LoTSS and FIRST peak flux densities and y-axis shows $\alpha^{1400}_{3000}$ derived using FIRST and VLASS peak flux densities. Regions in the plot for peaked, non-peaked and convex spectrum sources are marked with blue, orange and grey colour respectively. The spectral shape of a source in each quadrant is drawn at the corners. The error bar shows the typical error on the spectral indices. The hatched design marks the area which we classify as flat-spectrum region. Sources lying in this region are not included in peaked, non-peaked or convex spectrum groups. \OIII~detections are marked with filled circles, and non-detections are marked with hollow circles.}
\label{colour_colour_plot}
\end{figure}

Our aim is to link the AGN life cycle to feedback. The AGN life cycle can be probed using the radio spectral shape. To characterise the spectral shape, we have plotted our sources on a colour-colour plot of spectral indices, from 144-1400\,MHz and 1400-3000\,MHz, shown in Fig.~\ref{colour_colour_plot}. Depending on the location of a source on this plot, we can identify its radio spectral shape. As mentioned in the introduction, in the very early stages of evolution, the radio spectra of a radio AGN is peaked at GHz frequencies. As the source evolves, the spectral peak moves to lower frequencies and both optically thick and thin parts of the spectra can be observed. Ultimately, the radio source becomes completely optically thin and the spectral peak cannot be detected anymore. With our wide frequency coverage from 144\,MHz to 3000\,MHz, we can track the spectral peak over a large range, and identify sources in different stages of evolution.\par

Sources with $\alpha^{144}_{1400}$ > 0 \& $\alpha^{1400}_{3000}$ > 0 have an absorbed spectra with a spectral peak above 3000\,MHz. These sources are in the 'youngest' stage of evolution. As the spectral peak moves to lower frequencies, $\alpha^{1400}_{3000}$ becomes optically thin (< 0) while $\alpha^{144}_{1400}$ is still optically thick (> 0). These are GPS and CSS sources. They lie in the blue region of Fig.~\ref{colour_colour_plot} and we label them collectively as peaked spectrum sources. Although GPS and CSS sources are usually defined to have a high frequency spectral index $\alpha\leq-0.5$ \citep{ODea1998}, this definition is arbitrary and the high frequency spectral index has been observed to have a continuous distribution \citep{Fanti1990,Stanghellini1998,Callingham2017}. \par
When the spectral peak moves to frequencies below 144\,MHz, that is  $\alpha^{144}_{1400}$ < 0 \& $\alpha^{1400}_{3000}$ < 0, the radio spectra is completely optically thin in our frequency range. These are the most evolved sources in our sample. They lie in the orange region of Fig.~\ref{colour_colour_plot} and we label them non-peaked sources. It is possible that there are still some sources in this region that have a peak between 144\,MHz and 1400\,MHz, and more frequency points are needed to identify them. Sources in the grey region have a convex spectrum, that is  $\alpha^{144}_{1400}$ < 0 \& $\alpha^{1400}_{3000}$ > 0, and could be sources with multiple epochs of activity, where the optically thick high frequency spectral index is due to the newer phase of activity.\par

In Fig.~\ref{colour_colour_plot}, sources within the hatched region are not included in any spectral shape group. Sources in this region can have a flat spectra either due to the core dominating the emission, or relativistic beaming due to a small jet viewing angle. These sources are called flat-spectrum radio AGN, and are usually defined in literature to have a spectral index $\alpha \leq 0.5$ at all frequencies \citep{ODea1998}. In the plot, this region is within spectral indices $-0.5\geq\alpha^{144}_{1400}\leq0.05$ and $-0.5\geq\alpha^{1400}_{3000}\leq0.15$. This includes sources with spectral indices within 1$\sigma$ of both axes. It is also possible that there are some genuine peaked spectrum sources in this region, but we cannot identify them with the current frequency points. All spectral shape groups and their properties are summarised in Table~\ref{sample_properties}.\par

Although peaked spectrum sources are understood to be young radio AGN, another explanation used for these sources is the frustration hypothesis. In this case, peaked sources are not young AGN but confined to small scales by a dense ambient medium (for example,  \citealt{VanBreugel1984a,ODea1991,Callingham2015a}). Such a dense ambient medium can cause absorption of the radio emission by FFA. Therefore, we also check further in Section~\ref{radiospectralshaperesult} if the peaked sources in our sample are genuine young sources, or absorbed due to a high density medium. 

\subsubsection{Radio morphology and power}
\label{radio morphology}

The sample has been classified into compact and extended radio sources by \citetalias{Gereb2014}, using a combination of 1.4\,GHz NRAO VLA Sky Survey (NVSS, \citealt{Condon1998}) major-to-minor axis ratio and FIRST peak-to-integrated flux ratio (see section 2.1 of their paper). However, the low frequency and high sensitivity of LoTSS to diffuse low surface brightness emission means that it can detect low level extended emission, which could be missed at 1.4\,GHz. Therefore, we used the criteria  of \citet{Shimwell2022} for LoTSS-DR2, to classify a source as resolved or unresolved. Their criteria (described in detail in Section 3.1 of their paper) uses a combination of signal-to-noise ratio and integrated-to-peak flux ratio to separate sources. This method attempts to account for calibration inaccuracies and signal-to-noise ratio of the sources, and is more robust than just using the catalogued source sizes. However, it is important to note that the LoTSS criteria was designed for a statistically large sample, and can have field-to-field variations leading to a small percentage of sources being misclassified.\par
At the average redshift of our sample, an unresolved source with 6\arcsec resolution of LoTSS corresponds to a physical size upper limit of $\sim$12\,kpc. This is close to the typical upper limit of $\sim$20\,kpc for CSS sources (for example, \citealt{ODea1998,Fanti1990}). Therefore, we consider a source to be compact only if it is classified as unresolved by \citet{Shimwell2022} and compact by \citetalias{Gereb2014} criteria. All other sources are classified as extended. A visual inspection of the compact sources revealed that eight sources show clear extended emission in their LoTSS images, on scales of a few tens to hundred kpc. This could be due to the caveat in using the LoTSS criteria mentioned in the paragraph above. However, these source do not show any extended emission in their FIRST or VLASS images. Therefore, these sources could be restarted radio AGN, as mentioned in the introduction, and we discuss them in section~\ref{restarted section}. These sources were also moved to the extended group and are shown in Fig.~\ref{new_extended}. In summary, our sample has 64 compact and 65 extended radio sources. Out of the peaked sources discussed in the previous section, 68\% are compact and 32\% are extended. Comparing this to the non-peaked sources, where 52\% are compact and 48\% are extended, shows that peaked sources tend to be more compact than non-peaked sources, as would be expected from the AGN life cycle scenario.\par

We also calculated the 1.4\,GHz radio power of the sources using their peak flux density from FIRST. The distribution of the radio powers is shown in Fig.~\ref{sample_properties_distribution}. To assess the impact of radio power on feedback, we split the sample along the median radio power. The groups are summarised in Table~\ref{sample_properties}.

\subsection{Optical properties}
\subsubsection{Stellar continuum and emission line modelling}
\label{data reduction}
The stellar continuum and emission line modelling (using an F-test approach) presented for the sample was done in \citet{Santoro2018b}. In this section, we describe the procedure to obtain the emission line spectra of the sources. We have taken the optical spectroscopic data from SDSS DR12 \citep{Alam2015}. The spectra cover the wavelength range 3800-9200\,\AA~with an average spectral resolution of about 160\,\kms. The accuracy on the wavelength calibration is <5\,\kms. The SDSS pipeline provides properties of the emission lines, but only with a single-component fit to the line profile. A first visual inspection of the spectra already indicated that, when emission lines were present, they often had complex line profiles which required multiple kinematical components for proper modelling. For this reason we re-processed the SDSS spectra with our fitting procedure to extract more reliable information on the kinematics and excitation of the ionised gas. 

Considering that in a number of sources the emission lines are weak and the stellar continuum dominates, we performed a detailed fitting of the stellar continuum. The modelling was done using pPXF \citep{Cappellari2017} in combination with single stellar population models (SSP) from the 2016 version of the spectral synthesis results of \citet{Bruzual2003}. We used stellar populations with old (11 and 12.3\,Gyr), intermediate (0.29, 0.64, 0.90, 1.4, 2.5 and 5 Gyr) and young (5, 25 and 100 Myr) ages. All the SSP assume a Chabrier initial mass function, solar metallicity and instantaneous starburst. We shifted the spectra to the restframe using the SDSS redshift estimate and, during the fitting procedure, we masked the spectral regions corresponding to the \OII$\mathrm{\lambdaup\lambdaup}$3726,28\AA, \hdelta, \hgamma, \hbeta, \OIII$\mathrm{\lambdaup\lambdaup}$4958,5007\AA, \OI$\mathrm{\lambdaup}$6300\AA, \Nii $\mathrm{\lambdaup\lambdaup}$6548,84\AA, \halpha~and \Sii$\mathrm{\lambdaup\lambdaup}$6717,30\AA~emission lines. This procedure reproduces the stellar continuum of our sources well, including the cases where strong emission lines are present. An example of a source spectrum and stellar continuum modelling result is shown in Fig.~\ref{stellarcontinuum}.\par

After subtracting the best model of the stellar continuum given by pPXF, we performed the modelling of the emission lines using Gaussian functions.  
As a reference model, we used the \OIII$\mathrm{\lambdaup\lambdaup}$4958,5007\AA~doublet which is usually the strongest spectral feature associated with the ionised gas and is not contaminated by other emission lines. Each kinematical component of the model consisted of two Gaussian functions with the same width (that is  $\sigma_\mathrm{[OIII] 4958}=\sigma_\mathrm{[OIII] 5007}$), fixed separation and relative amplitude ($A_\mathrm{[OIII] 4958}=\frac{1}{3}\times A_\mathrm{[OIII] 5007}$) according to atomic physics. We started by fitting each line of the doublet with one Gaussian component and added up to two extra Gaussian components to allow our procedure to fit complex line profiles with broad wings. Every time we added a Gaussian component in our fitting procedure, any change in $\chi^{2}$ was evaluated with an F-Test to establish whether this provided a better fit at a confidence level >99\%. Multiple Gaussian components were required to fit the spectra of 21 sources. We considered the \OIII$\mathrm{\lambdaup}$5007\AA~emission line detected when the peak of the emission line model was higher than three times the standard deviation of the model residuals. If the \OIII$\mathrm{\lambdaup}$5007\AA~line was detected, we proceeded with fitting the other emission lines. \par
The best model of the \OIII~line was used to fit the \hbeta~line, that is  up to three Gaussian components per emission line. To take into account possible \hbeta~broad emission coming from the broad line region (BLR) of type 1 AGN, we allowed one more Gaussian component in the fit of the \hbeta~line if needed. In our sample, two sources required an additional broad component, and are classified as BLR. Finally, we fit the \Nii$\mathrm{\lambdaup\lambdaup}$6548,84\AA~and the \halpha~lines together. Each kinematical component of this model was made by three Gaussian functions with fixed separation according to atomic physics and the same line width. We also fixed the relative line ratio of the two Gaussian functions associated with the \Nii~lines ($A_\mathrm{\Nii 5648}=\frac{1}{3}\times A_\mathrm{\Nii 5684}$). Both emission lines were fit using the best model for the \OIII~line. We included an extra Gaussian component to fit the \halpha~line when an additional component was needed to fit the \hbeta~line. \par

We extracted the fluxes of the \OIII, \halpha, \hbeta~and \Nii~emission lines considering the summed flux of all the components needed to model the line. If an additional component is needed to fit the \hbeta~and \halpha~line, we extracted the flux of these lines only from the components of the \OIII~model. In this way our line fluxes were not affected by the light coming from the BLR of the AGN. Overall, in the spectra of the 129 radio AGN, we have 93 \OIII~detections. For the non-detections, we estimated an upper limit on the flux and used it to estimate the upper limit for the \OIII~luminosity. The luminosity distribution is shown in Fig.~\ref{sample_properties_distribution}. In this radio AGN sample, we have 77 \halpha~detections, 55 \hbeta~detections and 91\Nii~detections. Fifty five sources have detections for all four lines and 21 sources have detections for \OIII, \halpha, and \Nii~but not \hbeta. Fifteen sources have detections for \OIII, and \Nii~but not \halpha~and \hbeta. One source has detection for \OIII, and \halpha~but not \hbeta~and \Nii. One source has only a detection for \OIII~and no other line. \par
The \OIII~luminosity is also a commonly used proxy for the total radiative luminosity of an AGN. We used it as a proxy for assessing the role of radiation pressure in driving the gas kinematics. To do this, we split the sample about the median \OIII~luminosity, $L_\mathrm{[OIII]}=1.6\times10^{40}$\,\ergsunit (including the upper limits for non-detections). The groups are summarised in Table~\ref{sample_properties} and the results shown in Section~\ref{optical classification}. We did not use the extinction corrected \OIII~luminosity because there were very few sources, only 55, with both \halpha~and \hbeta~detections. Thus the correction could only be performed for part of the sample. Furthermore, the relation to convert \OIII~luminosity to the total radiative luminosity, discussed in Section~\ref{eddington ratios}, uses the observed \OIII~luminosity. We note that although we have used the \OIII~luminosity as a proxy for the radiative luminosity, for our sample of radio AGN, the \OIII~emission is not solely due to AGN photoionisation, but can also have contribution from shock ionisation. This could affect our result for disentangling the role of radiation pressure and is a caveat of this method. \par

Since the \OIII~line modelling requires multiple Gaussian components in some cases, a comparison between them is hard. To compare the kinematics of the individual \OIII~detections in a uniform manner, following the approach of \citet{Mullaney2013}, we estimated the flux-weighted average FWHM of the \OIII$\lambdaup5007$\AA~line (FWHM$_\textrm{avg}$), defined as 
\begin{ceqn}
\begin{align}
      \textrm{FWHM}_\textrm{avg} = \sum\limits_{i}\textrm{F}(i) \times \textrm{FWHM}(i),
\end{align}
\end{ceqn}
where F$(i)$ is the flux ratio of the $i$th component to the total flux in the model, and FWHM$(i)$ is the full width at half maximum of the $i$th component. This allows for a direct comparison between the different \OIII~profiles, regardless of the number of Gaussian functions needed to model the line.\par 

Multi-component model fits can help track non-gravitational motion in the NLR, for example, a gas outflow. For instance, if they have a significantly different velocity than the systemic (core component), or broader widths than expected from gravitational motion. Throughout the paper, whenever a multi-component fit is required to fit the \OIII~spectra, we used two criteria to identify the extra ("wing") component as a gas outflow. These are - (i) the velocity difference from the core component, $\Delta v$, is at least $3\sigma$ significance, or (ii) the component's velocity dispersion, $\sigma_\textrm{{disp}}$, is larger than the three sigma upper limit of the stellar velocity dispersion. For the stellar velocity dispersion upper limit, we used the average value for the sample, which is 348\,\kms. We only used wing components whose amplitude is of at least 5$\sigma$ significance, where $\sigma_\mathrm{RMS}$ is the RMS noise in the line free region.

\subsubsection{BPT diagram}
\label{bpt diagram}
\begin{figure}
\includegraphics[width=\columnwidth]{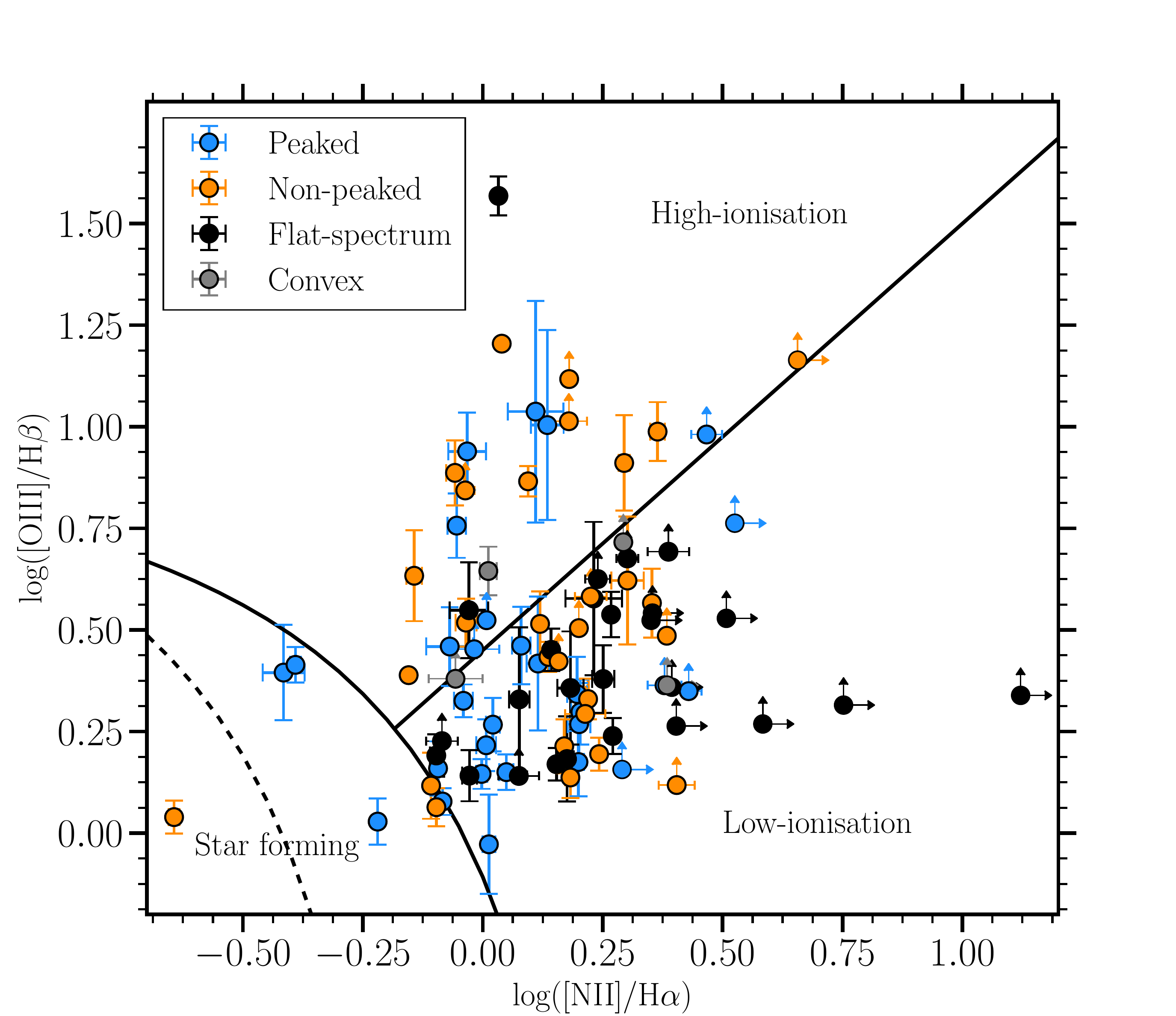}
\caption{ BPT diagram of the sources, with sources marked according to their radio spectral shape. The solid curve is the maximum starburst line from \citet{Kewley2001}. The dashed curve is the semi-empirical line from \citet{Kauffmann2003}. The solid line separating high and low ionisation sources is the empirical relation from \cite{Schawinski2007}.}
\label{bpt}
\end{figure}
To assess the role of radiation pressure in driving outflows in our sample, we also used the ionisation state of the sources. For this we used the BPT diagnostic diagram \citep{Baldwin1981, Kewley2006}. The BPT diagram uses a combination of ionised line flux ratios to identify whether the main source of ionisation is the light from the AGN or star formation. Using the emission line fluxes obtained in Sect.~\ref{data reduction}, we can plot our sources on the BPT diagram. We used the \halpha, \hbeta, \Nii, and \OIII~line flux ratios and require that at least one of \Nii~and \halpha, and one of \OIII~and \hbeta~are detected. This is needed in order to place the source on the BPT plot, even if its only an upper or lower limit. This gives us a total of 89 radio AGN, plotted in Fig.~\ref{bpt}.\par
We find that most of our sources lie in the AGN ionisation region (Seyfert or LINER), which is in agreement with the results of \citetalias{Maccagni2017a}, who also found that most of the sources lie above the MIR-radio relation, suggesting that the bulk of their radio emission comes from the central AGN rather than star formation. This confirms that the contribution from star formation to the radio emission and ionisation of the ISM, is insignificant in our sample. We then grouped the sources into high and low ionisation states, using the empirical relation from \citet{Schawinski2007}. Sources lying in the Seyfert region in Fig.~\ref{bpt} are classified as high ionisation sources and those in the LINER region are classified as low ionisation sources. We classified 23 sources as high ionisation and 58 sources as low ionisation (see Table~\ref{sample_properties}). Their \OIII~spectra is discussed in Sect.~\ref{optical classification}.\par

\subsubsection{Individual outflow detections}
\label{individual detections}
During our modelling analysis in Section~\ref{data reduction}, 21 sources required a multi-component model to fit the \OIII~spectra. Out of these, eighteen sources required double and three sources required triple kinematic components. As mentioned, multi-component fits hint towards the presence of kinematically disturbed and possibly outflowing gas. Using our aforementioned criteria, out of the 21 sources that had a multi-component fit, we identify 18 sources to have an outflowing component. This gives an outflow detection rate of $\sim$14\% for our sample. We discuss these sources in Section~\ref{oiiioutflows}.\par

Since we want to investigate the presence of outflows in sub-groups of sources and the number of individual outflow detections in our sample is relatively low, and we decided to use a stacking analysis for the rest of the paper. This allows us to explore, on average, the link between the gas kinematics and source properties. The stacking analysis is explained in the next  section, and the results are shown in Section~\ref{stacking results}. 

\subsubsection{Stacking procedure}
\label{stacked spectra}
As mentioned above, we detect individual outflows in a small number ($\sim$14\%) of sources. Studies of ionised gas outflows often rely on individual detections of the outflowing gas component, and this can bias them  towards the strongest outflows.  But since we aim to study the presence and dependence of outflows on the properties of radio AGN, it is important to include weak outflows that might not be detected individually with the sensitivity of the SDSS spectra. To do this, we stack the stellar continuum subtracted SDSS spectra, in groups of sources based on their radio and optical properties, to follow the evolution of the ionised gas kinematics.\par

We calculated the stacked spectra using a weighted average of the individual source spectra, defined as: 
\begin{ceqn}
\begin{align}
      S(\lambdaup) = \frac{\sum\limits_{i}\frac{1}{\sigma^{2}(i)}S(i,\lambdaup)}{\sum\limits_{i}\frac{1}{\sigma^{2}(i)}},
\end{align}
\end{ceqn}

where S($\lambdaup$) is the stacked spectrum, S$(i,\lambdaup)$ is the individual source spectra, and $\sigma(i)$ is the noise in a line free region of the source spectra. In the stacked spectrum, the noise is expected to go down by 1/$\sqrt{N}$, where N is the number of stacked sources. In our sample, stacking all the 129 radio AGN with an average noise of 1.33\ergs, gives a noise $\sim$0.19\ergs in the stacked spectrum. This is in agreement with the expected theoretical decrease by about a factor of 11. The errors on the stacked spectra were estimated using the bootstrapping method. We randomly selected 5000 samples from each group with replacement and stacked them. Using the 5000 stacked spectra, we estimated the errors at each point by taking the standard deviation of the stacked values at that point. The stacked spectra for various groups of sources are shown and discussed in Section~\ref{stacking results}.

\subsubsection{Stacked emission line fitting procedure}
\label{model fit}

For the analysis of the stacked spectra, we have used a Bayesian inference model fitting routine to assess the quality of the fit that best describes the profile. This is a probabilistic approach, and different from our method in section~\ref{data reduction}. It is better suited for assessing the average properties of a population of sources than the least squares approach. The method uses the Bayes theorem to obtain posterior probability density functions for the model parameters. In this routine, we estimate a likelihood function $\mathcal{L}(\theta)\equiv P(\Vec{D}|\Vec{\theta},M)$, which gives the probability of observing the data given the model parameters. We then used the Markov chain Monte Carlo methods to sample the posterior probability density functions of the model parameters such that this likelihood function was maximised. To implement this, we used the \texttt{emcee} package \citep{Foreman-Mackey2013}. For the \OIII~doublet, our model was made up of Gaussians with fixed width, mean, and relative amplitudes of the doublet, as described in Section~\ref{data reduction}. To account for a non-zero baseline in the spectra, we also added a first order polynomial of the form $a_\textrm{p}x+b_\textrm{p}$ to the model.\par

To assess the quality of the fit we used the Bayesian information criterion (BIC; \citealt{Schwarz1978}) which is estimated using a combination of the number of free parameters in the model, the number of data points and the likelihood value. This criteria is designed to give a lower value for a better fit, while punishing for adding free parameters. Therefore, the model with the least BIC is preferred. We define $\Delta \mathrm{BIC}=\mathrm{BIC}_{2}-\mathrm{BIC}_{1}$, where BIC$_{2}$ and BIC$_{1}$ are the values for the models with higher and lower number of parameters, respectively. The general rule of thumb is this: a -2$\leq$$\Delta$BIC<0 means there is weak evidence that adding more parameters improves the model fit, -6$\leq$$\Delta$BIC<-2 means there is moderate evidence that adding more parameters improves the fit, and -10$\leq$$\Delta$BIC<-6 means there is strong evidence, and $\Delta$BIC<-10 means there is very strong evidence that adding more parameters improves the fit \citep{Jeffreys1961}. For our analysis, we take a conservative approach and only consider a model with more parameters to improve the fit if $\Delta$BIC\,$\leq$\,-6.\par    

For the fitting of the stacked line presented below in Section~\ref{stacking results}, we started by fitting each line in the doublet with a single Gaussian, and performing this fit over 1000 times. The best fit parameters and the errors were then estimated from the average of the 1000 runs. These values were used to estimate the BIC. We then added another Gaussian component to the model and repeated the procedure. The best fit model was then picked depending on the change in BIC. We found that each line can be fit with one to two Gaussians. Fitting three Gaussians always worsened the fit quality, which is consistent with the fact that only three individual \OIII~detections required triple components. We apply this to the analysis of the stacked spectra of the various sub-groups as described in section~\ref{stacking results}. The results are summarised in Table~\ref{model fit} and best fit models are shown in Fig.~\ref{radio_spectralshape} and Fig.~\ref{modelfits_spectra}. 

\begin{figure}
\includegraphics[width=\columnwidth]{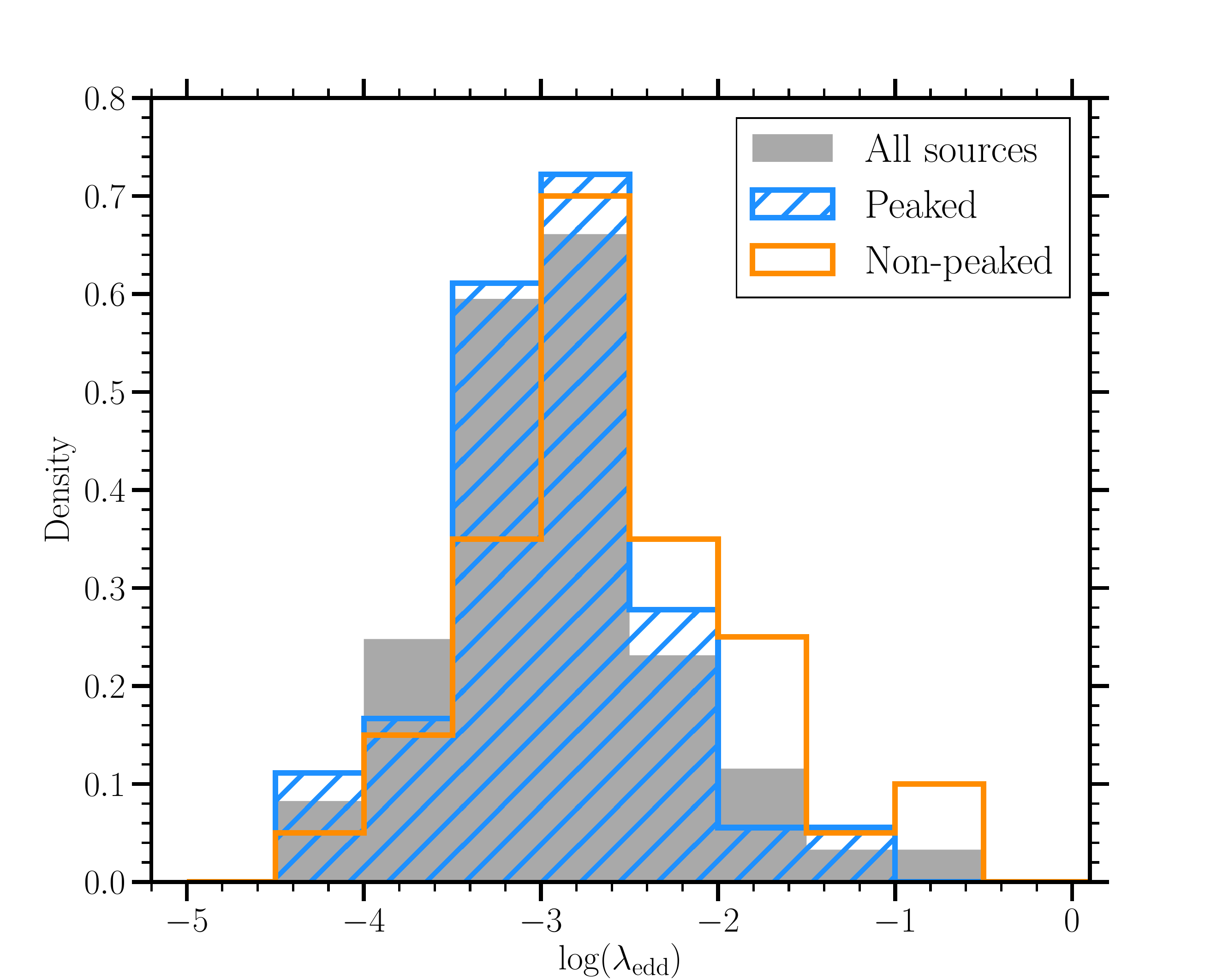}
\caption{Eddington ratio distribution of the 121 radio AGN with reliable estimates of stellar velocity dispersion, shown in grey. The distributions for peaked and non-peaked radio AGN are shown in blue and orange respectively. Both groups have a similar range of Eddington ratios. }
\label{eddratio distribution}
\end{figure}

\subsection{Eddington ratios}
\label{eddington ratios} 
Studies of ionised gas kinematics in AGN have found an increase in velocity dispersion and outflow detection rates with the Eddington ratio (for example, \citealt{Mullaney2013,Woo2016,Singha2021}). They conclude that this positive correlation is evidence for gas kinematics and outflows being driven by radiation pressure. To investigate this link, we calculate the Eddington ratio for our sources. The Eddington ratio can be calculated as the ratio of the total radiative output to the maximum luminosity that can be emitted by spherical accretion of gas on to the central SMBH, that is  the Eddington luminosity \citep{Best2012}. The Eddington ratio is then defined as $\lambdaup_\mathrm{Edd}=L_\mathrm{rad}/L_\mathrm{Edd}$.\par

For the radiative luminosity of our sources, we have used the relation of \citet{Heckman2004}, $L_\mathrm{rad}=3500\times L_\mathrm{[OIII]}$, where $L_\mathrm{[OIII]}$ is the observed \OIII~luminosity in the SDSS spectra. The scatter in this relation is $\approx$0.4\,dex. In order to determine the Eddington ratios for all sources, we have used the upper limit in luminosity for \OIII~non-detections. There is a caveat to using \OIII~luminosity as a proxy for the radiative luminosity for our sample, that we mentioned in Section~\ref{data reduction}. The Eddington luminosities for AGNs can be determined using the relation $L_\mathrm{Edd}=1.3\times10^{31} M_\mathrm{BH}/M_\mathrm{\odot}$\,W, where $M_\mathrm{BH}$ is the mass of the central SMBH in units of solar mass. To obtain the SMBH mass, we used the $M-\sigma_{\star}$ relation from \citet{McConnell2013}, $\mathrm{log}(M_\mathrm{BH}/M_\mathrm{\odot})=8.32+5.64\,\mathrm{log}(\sigma_{\star}/200\,\mathrm{km\,s^{-1}})$, where $\sigma_{\star}$ is the stellar velocity dispersion. This relation has a scatter of $\approx$0.4\,dex. We used the stellar velocity dispersion measurements from SDSS. Velocity dispersion values $\leq100$\,\kms and $\geq850$\,\kms were unreliable and have been removed. Finally, we have estimated the Eddington ratios for 121 sources in our sample. The distribution is shown in Fig.~\ref{eddratio distribution}. We split the sample about the median Eddington ratio to assess its role in feedback, and show the results in Section~\ref{eddingtion ratio classification}.

\section{Results}
\label{results}
The goal of this study is to connect the properties of the AGN with the one of the ionised gas. We not only consider the radio and optical properties, but also focus on the group we have identified as young radio AGN. As mentioned in the introduction, some trends with these properties have been found but, with all the parameters derived for our sample and discussed in section~\ref{methods}, we are now able to assess their role, and use stacking a stacking analysis to disentangle the role of these properties. We start by discussing the individual outflow detections we have obtained, followed by the stacking results.

\subsection{\OIII~outflows}
\label{oiiioutflows}

\begin{figure}
\includegraphics[width=\columnwidth]{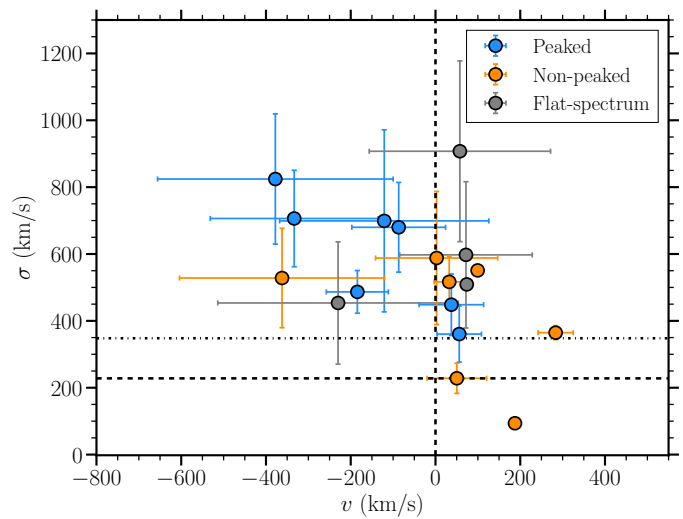}
\caption{Velocity (with respect to systemic velocity) vs dispersion of the wing components identified as outflows. Sources are coloured according to their radio spectral shape. Horizontal lines show the average stellar velocity dispersion of the sample and its three sigma upper limit. Peaked spectrum sources show broader outflows with larger velocities, compared to non-peaked sources.}
\label{mom1mom2}
\end{figure}

As mentioned in Section~\ref{individual detections}, 18 sources have an outflowing component in their \OIII~spectra according to our criteria. These components are shown in Fig.~\ref{mom1mom2} which plots their velocity difference from the core component vs their velocity dispersion. We find that the peaked sources have broader outflowing components, typically with a higher velocity dispersion than their non-peaked counterparts. This suggests that the kinematics of the ionised gas is linked to the radio spectral shape. Physically, this hints towards more disturbed gas at the younger stages of radio AGN life cycle. Although we see hint of a trend with the radio spectral shape in Fig.~\ref{mom1mom2}, the low number of individual outflow detections ($\sim$14\%) makes it hard to draw a link with the AGN life cycle. Therefore, to explore this trend for the entire sample, we stacked the optical spectra for all the sources, as mentioned in Section~\ref{stacked spectra}. Below we show the results from the stacking analysis.

\subsection{Stacking results}
\label{stacking results}
\subsubsection{Radio spectral shape classification}
\label{radiospectralshaperesult}
\begin{figure*}
  \includegraphics[width=2\columnwidth]{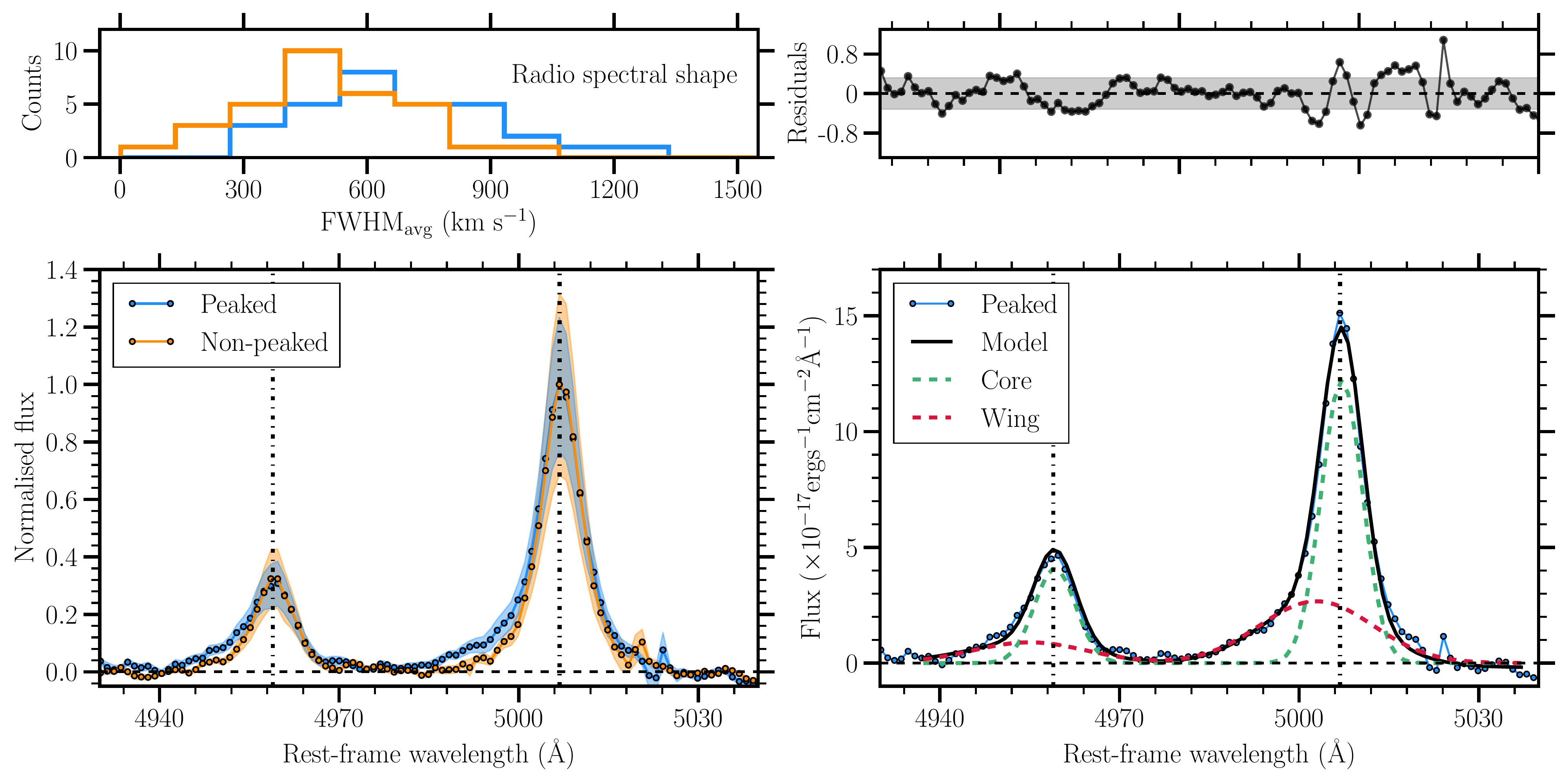}
  \caption{Stacking results for peaked and non-peaked sources. \textit{Top left.} Average FWHM distributions of the \OIII~detections in the peaked and non-peaked groups. \textit{Bottom left.} Stacked \OIII~spectra of peaked and non-peaked sources, including non-detections. The shaded regions show the errors in the stacked spectra. The peaked sources show a significantly broad spectra on the blueshifted side, than the non-peaked sources. \textit{Bottom right.} Model fit to the peaked spectra, with the core and wing components marked. \textit{Top right.} Residuals after subtracting the model from the peaked spectra. The shaded region marks the standard deviation of the residuals. }
  \label{radio_spectralshape}    
\end{figure*}

\begin{table*}
\caption{Model fit parameters}
         \label{model fit}
\renewcommand{\arraystretch}{1.05}
\setlength{\tabcolsep}{4pt}
\begin{tabular}{cccccccccc}
            \hlineB{4.5}
            
    \noalign{\smallskip}
     Groups & $A_\textrm{core}$  & FWHM$_\textrm{core}$ & $v_\textrm{core}$ & $A_\textrm{wing}$ &  FWHM$_\textrm{wing}$ &  $v_\textrm{wing}$ & $a_\textrm{p}$ & $b_\textrm{p}$ & $\Delta$BIC \\
     & [$\times10^{-17}$] &  &  & [$\times10^{-17}$] &   &  & [$\times10^{-3}$] & [$\times10^{-1}$] & \\
    \noalign{\smallskip}
    \hlineB{4.5}
    \noalign{\smallskip}

    \textit{Radio spectral shape} \\
    Peaked & 12.2$\pm$0.9 & 486$\pm$48 & 18$\pm$11 & 2.7$\pm$0.8 & 1339$\pm$418 & -237$\pm$102 & -2.5$\pm$4.4 & -1.1$\pm$2.1 & -11 \\
    Non-peaked & 13.8$\pm$0.9 & 521$\pm$46 & 21$\pm$14 & - & - & - & 4.7$\pm$16.0 & 1.2$\pm$1.2 & 5 \\
    \noalign{\smallskip}
    \hline
    \noalign{\smallskip}
    \textit{Radio morphology} \\
    Compact & 15.2$\pm$0.5 & 571$\pm$24 & 17$\pm$8 & - & - & - & -0.6$\pm$2.3 & 1.2$\pm$1.3 & 10 \\
    
    Extended & 6.9$\pm$0.5 & 597$\pm$59 & 3$\pm$19 & - & - & - & -2.2$\pm$2.5 & 0.004$\pm$1.2 & 16 \\    
    \noalign{\smallskip}    
    \hline
    \noalign{\smallskip}
    \textit{Central 1.4\,GHz luminosity} \\
    Low luminosity & 7.9$\pm$0.7 & 602$\pm$70 & 8$\pm$23 & - & - & - & 2.6$\pm$15.0 & 0.6$\pm$1.2 & 20 \\
    High luminosity & 14.9$\pm$0.5 & 562$\pm$25 & 23$\pm$8 & - & - & - & -1.7$\pm$2.5 & 0.7$\pm$1.2 & 1 \\
    \noalign{\smallskip}    
    \hline
    \noalign{\smallskip}    
    \textit{BPT diagram} \\
    Low-ionisation  & 9.1$\pm$0.5 & 662$\pm$45 & -5$\pm$16 & - & - & - & -3.2$\pm$2.7 & -0.5$\pm$1.2 & 17 \\
    High-ionisation & 32.1$\pm$1.9 & 450$\pm$33 & 28$\pm$4 & 8.4$\pm$2.0 & 1219$\pm$143 & 0$\pm$26 & -1.8$\pm$2.7 & -0.7$\pm$1.4 & -93 \\
    \noalign{\smallskip}    
    \hline    
    \noalign{\smallskip}    
    \textit{Optical luminosity} \\
    Low luminosity & 4.4$\pm$0.5 & 582$\pm$88 & -3$\pm$29 & - & - & - & -2.1$\pm$2.5 & -0.8$\pm$1.2 & 16 \\
    
    High luminosity & 17.0$\pm$0.5 & 585$\pm$23 & 17$\pm$7 & - & - & - & -0.6$\pm$2.3 & 1.8$\pm$1.4 & -2 \\
    \noalign{\smallskip}
    \hline
    \noalign{\smallskip}    
    \textit{Eddington ratio} \\
    Low ratio & 4.3$\pm$0.5 & 622$\pm$111 & 0$\pm$29 & - & - & - & -3.8$\pm$10.0 & -0.8$\pm$2.6 & 12 \\    
   
    High ratio & 14.5$\pm$0.5 & 548$\pm$33 & 6$\pm$9 & - & - & - & 0.6$\pm$10.0 & 1.6$\pm$1.3 & 2 \\
    \noalign{\smallskip}    
    \hlineB{4.5}
 \end{tabular}
    \flushleft
    \textbf{Note.} Model fit parameters for the stacked \OIII~spectra in the different groups. $A_\textrm{core/wing}$ are the amplitudes in $\mathrm{erg s^{-1} cm^{-2} \AA^{-1}}$. FWHM$_\textrm{core/wing}$ are the full widths at half-maximum in \kms. $v_\textrm{core/wing}$ are the velocities of the components with respect to the systemic velocity in \kms. All values are for the $\lambdaup5007$\,\AA~line. $a_\mathrm{p}$ and $b_\mathrm{p}$ are coefficients of the linear polynomial. The model parameters are described in Section~\ref{model fit}. $\Delta$BIC is the information criteria difference between the double and single component model fits, that is  BIC$_\mathrm{double}$-BIC$_\mathrm{single}$.
    
\end{table*}

\begin{figure*}
  \includegraphics[width=2\columnwidth]{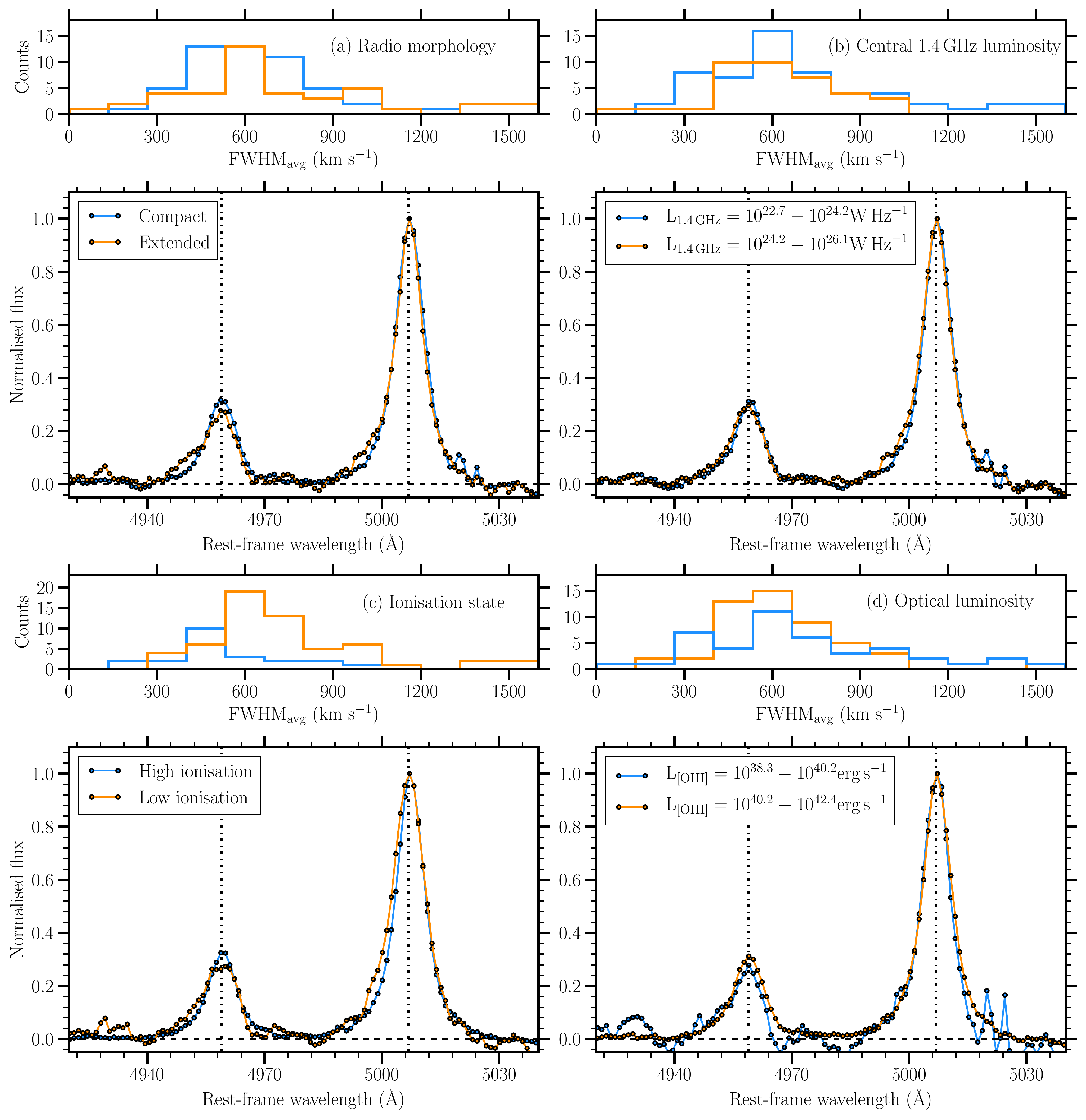}
  \caption{Stacked \OIII~spectra and FWHM$_\mathrm{avg}$ distributions for groups based on (a) radio morphology, (b) central 1.4\,GHz luminosity, (c) BPT diagram and (d) optical luminosity. The top panel in each sub figure shows the distribution of the \OIII~detections in the groups. Bottom panel in the sub figures shows the stacked \OIII~spectra (including non-detections). The fluxes have been normalised to the maximum value for easier comparison of the profile shapes. We only see a significant difference in the stacked spectra based on BPT classification (c).}
  \label{stackedspectra_groups}    
\end{figure*}
The stacked \OIII~spectra and model fits for the peaked and non-peaked sources are shown in Fig.~\ref{radio_spectralshape}. The peaked sources clearly show a broader spectra on the blueshifted side than the non-peaked sources. The FWHM$_\mathrm{avg}$ distributions of the two groups also show a higher velocity tail for the peaked sources. A two sample KS test gives a statistic of 0.34 and a p value of 0.04, which suggests that the two distributions are significantly different at 95\% confidence level. This confirms our result in the previous section, that the \OIII~kinematics is linked to the radio spectral shape. To check whether this difference in stacked spectra is due to the individual outflow detections, we re-performed the stacking analysis after removing the outflow detections. We found that the stacked spectra for peaked sources still had a broad component on the blueshifted side, compared to the non-peaked sources. However, the difference is not as strong as before, due to the decrease in the number of sources stacked. A two sample KS test for these groups' distribution of \OIII~average FWHM gives a statistic of 0.34 and a p value of 0.09, which suggests that the two distributions are significantly different at 90\% confidence level. This confirms the fact that our result for peaked and non-peaked sources is true on average for sources, and is not affected by individual outflow detections.\par

Our modelling analysis shows that a two component model provides the best fit for the peaked sources, whereas a single component model gives the best fit for the non-peaked sources (see Table~\ref{model fit} and Fig.~\ref{radio_spectralshape}). Using our criteria, it is clear that the peaked sources have a blueshifted outflow, whereas non-peaked sources do not. The outflowing component has a peak velocity of $-237\pm102$\,\kms, and is very turbulent, with a FWHM of $1330\pm418$\,\kms (after deconvolving velocity resolution). The outflow component also contains $\approx$38\% of the total \OIII~flux. This suggests that on average, the \OIII~gas kinematics evolves with the radio jets life cycle, and that the ionised gas is most disturbed when the radio spectra is peaked, that is  at the youngest stage of radio jet evolution.\par
As mentioned in Section~\ref{spectral classification}, peaked sources could also be explained by the "frustration" hypothesis due to a dense ambient medium. To estimate the electron densities of the NLR region, we used the stacked \Sii~doublet and the relation from \citet{Sanders2015}, to obtain electron density of $n_{e}=237\pm52\,\textrm{cm}^{-3}$ for peaked and $n_{e}=243\pm55\,\textrm{cm}^{-3}$ for non-peaked sources. We note that this relation between \Sii~ratio and $n_\textrm{e}$ saturates at a ratio of $\sim$0.44 and cannot be used for gas with $n_\textrm{e}>10^{5}\textrm{cm}^{-3}$. However, this should not affect our results because the electron density we obtain is significantly lower than this limit. We then estimate ionised gas masses of $M_\textrm{ion}=4.6\times10^{5}M_{\odot}$ for peaked, and $M_\textrm{ion}=6.4\times10^{5}M_{\odot}$ for non-peaked sources \citep{Revalski2022}. The electron densities and ionised gas masses are similar for the two groups. This shows that peaked sources typically do not have a denser ambient medium that non-peaked sources in our sample. We also calculate that for the radio spectra to have a peak in the GHz frequencies ($\gtrsim1.4$\,GHz), if surrounded by a medium of such electron densities, the path length would have to be $\gtrsim$\,10\,kpc. It is highly unlikely that the NLR clouds exist out to such large radial distances for majority of our sources.  Therefore, we are confident that our peaked sources, on average, are not "frustrated" or absorbed due to FFA, and are genuine young sources (which nevertheless appear to be interacting with the surrounding medium, likely slowing their expansion for example, \citealt{Bicknell2006,Wagner2010,Bicknell2018}). Some of our sources are also included in the \citet{DeVries2009} sample, who have used VLBI data to conclude that these are young radio AGN, absorbed due to SSA. 

\subsubsection{Radio morphology and luminosity}
\label{radio classification result}
To assess the impact of radio morphology and luminosity on the \OIII~gas kinematics, we also stacked the sources in groups of compact and extended morphology, and high and low 1.4\,GHz radio luminosity (see Table~\ref{sample_properties} for details). The stacked spectra are shown in the top panel of Fig.~\ref{stackedspectra_groups}. We find no significant difference in the stacked spectra based on groups of radio morphology or 1.4\,GHz luminosity. The modelling analysis also shows that the best fit is provided by a single component model (see Table~\ref{model fit} and Fig.~\ref{modelfits_spectra}). Although it can be seen in Fig.~\ref{modelfits_spectra} that there is some emission on the blue side that could be fit with another component, we found the improvement in the fit with a double component model was not significant enough to be considered the best fit model, according to our BIC criteria ($\Delta\textrm{BIC}\leq-6$). This does not mean that there are no outflows in either group. The main result is the fact that, based on the lack of a strong difference in the stacked profiles, combined with the model fit results, we do not find a strong dependence of the presence of outflows on these properties. A two sample KS test of the average FWHM distributions for compact and extended sources gives a statistic of 0.22 and a p value of 0.21. Similarly for radio luminosity, we obtain a ks statistic of 0.16 and a p value of 0.59. Thus we find no significant differences between the average FWHM of the \OIII~profiles in these groups. \par

To check the robustness of this result, we performed the same stacking analysis by changing our classification criteria of source morphology and using either only the criteria from \citet{Shimwell2022} or \citetalias{Gereb2015}. This did not affect our results. The lack of significant difference in the stacked spectra of radio AGN on the basis of radio morphology and luminosity shows that these are not the dominant properties affecting \OIII~gas kinematics in our sample, and the link between the gas kinematics and radio spectral shape.
\subsubsection{Ionisation ratio and \OIII~luminosity}
\label{optical classification}
Our final goal is to be able to disentangle the effects of radio properties from radiation in our sample. Therefore, to assess the role of radiation pressure in driving the \OIII~gas kinematics, we stacked the spectra in groups of two classifications - the ionisation state, using the BPT diagram (Section~\ref{bpt diagram}) and the observed \OIII~luminosity. The stacked spectra are shown in the bottom panel of Fig.~\ref{stackedspectra_groups}. Visual inspection suggests a difference in the width of the profiles as can be seen in the figure. Our modelling analysis (Fig.~\ref{modelfits_spectra}) also indicates a difference between high and low ionisation state groups as can be seen in Table~\ref{model fit}. The core component of the low ionisation group has a broader FWHM of $645\pm45$\,\kms (after deconvolution of the instrument resolution) than the high ionisation group FWHM of $425\pm33$\,\kms. However, the low ionisation group also has a larger average stellar velocity dispersion ($\sigma_{\star}=230\pm8$\,\kms) than the high ionisation group ($\sigma_{\star}=170\pm8$\,\kms). The \OIII~velocity dispersion of the core component has been known to correlate with the stellar velocity dispersion for Seyferts \citep{Greene2005a,Komossa2008}, and recently, such a trend has also been found for type II AGN \citet{Woo2016,Woo2017}. Therefore, this difference in the FWHM of the core component between the low and high ionisation groups can be related to the difference in the average stellar velocity dispersion. Interestingly, the high ionisation state profile is best fit with a double component model, whereas the low ionisation state profile is best fit with a single component model. The wing component in high ionisation group has a FWHM of $1208\pm143$\,\kms (after deconvolving velocity resolution), and is identified as an outflow according to our criteria. A two sample KS test of the average FWHM distributions between the two groups also gives a statistic of 0.45 and a p value of 0.002, suggesting a difference between the \OIII~profiles at >99\% confidence level. These results suggest a link between the ionisation state and the \OIII~kinematics. \par 

However, for the groups in \OIII~luminosity,
we find no significant difference in the stacked spectra (Fig.~\ref{stacked spectra}). Both are best fit with a single component model according to our criteria, although there is some emission on the blue side in Fig.~\ref{modelfits_spectra} that could be fit with another component. However the improvement is not significant enough to be considered as best fit model. Therefore, on average, we find no dependence of the presence of an outflow on \OIII~luminosity. Re-performing the analysis using only \OIII~detections also did not affect our results. A two sample KS test of the average FWHM distributions also gave a statistic of 0.18 and a p value of 0.44, thus showing no significant difference between the two groups. These results from the ionisation state and \OIII~luminosity groups appear contradictory in determining the role of radiation pressure in our sample. This could mean that radiation pressure does not play a definitive role in our sample of radio AGN, and we discuss it further in Section~\ref{other properties discussion}.

\subsubsection{Eddington ratio}
\label{eddingtion ratio classification}
\begin{figure}
\includegraphics[width=\columnwidth]{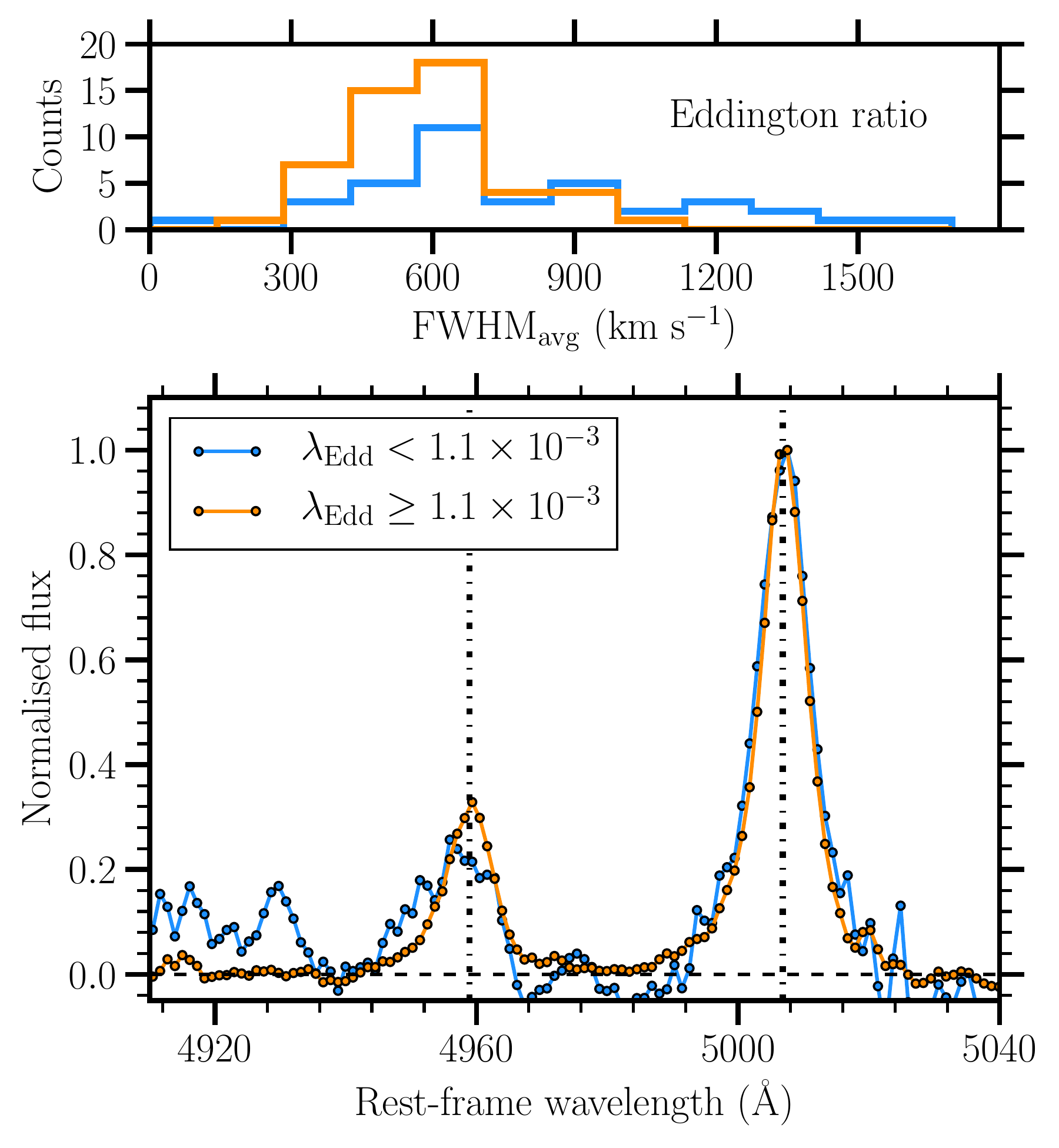}
\caption{Stacked results for Eddingtion ratio.  (\textit{Top}) Average FWHM distributions of the \OIII~detections and  (\textit{Bottom}) stacked \OIII~spectra for the high and low Eddington ratio groups. The fluxes have been normalised to the maximum value.}
\label{stacked_eddratio}
\end{figure}

As discussed in Section~\ref{eddington ratios}, we split our sample into two groups of Eddington ratios to investigate their link with \OIII~kinematics. The stacked spectra are shown in Fig.~\ref{stacked_eddratio} and show a good overlap with each other. The modelling analysis (see Table~\ref{model fit} and Fig.~\ref{modelfits_spectra}) shows that both spectra are best fit with a single component model. A two sample KS test of the average FWHM distributions gives a statistic of 0.34 and a p value of 0.01, suggesting a difference between the \OIII~profiles at >99\% confidence level. However this difference is likely due to the larger average stellar velocity dispersion of the lower Eddington ratio group, by $\approx$43\,\kms, as also discussed in Section~\ref{optical classification}.\par

\section{Discussion}
\label{discussion}
Our aim has been to characterise the evolutionary stage for a sample of radio AGN, and study feedback on the ionised gas over the life cycle of these AGN. By using a stacking analysis in groups based on the optical and radio properties of the sources, we have tried to disentangle their role in driving the ionised gas outflows. In this section, we discuss our results and the link between the life cycle and feedback on the ionised gas. We then estimate the outflow energetics, and present our conclusions. 

\subsection{Link to radio AGN life cycle}
\label{linktolife-cycle}
The main result of our paper is the link between radio spectral shape and the presence of outflow, shown in section~\ref{radiospectralshaperesult}. We find that stacking the peaked sources together shows a more prominent blueshifted component in the \OIII~profile, whereas stacking non-peaked sources does not (see Fig.~\ref{radio_spectralshape} and Table~\ref{model fit}). We interpret this as outflowing gas in the peaked sources. This outflow is very broad, with a  deconvolved FWHM = $1330\pm418$\,\kms. In section~\ref{oiiioutflows} and Fig.~\ref{mom1mom2}, we also find that individually detected outflows in peaked spectrum sources have higher velocities and widths than non-peaked sources. These results lead us to conclude that the feedback on the ionised gas is linked to the radio spectral shape of the sources. Since the spectral shape is a good indicator of the evolutionary stage, feedback is therefore linked to the life cycle of the AGN. This shows that impact of the jets evolves over the AGN life cycle. The ionised gas kinematics is most disturbed when the radio jets are young (peaked sources), and it becomes less extreme as the jets evolve (non-peaked sources). \par
The outflow velocity of $-237\pm102$\,\kms in peaked sources is lower than those found in powerful radio galaxies. For example, for a sample of 3C radio AGN, \citet{Speranza2021} found outflowing gas with blueshifted velocities ranging from $\sim$300\kms to $\sim$900\kms. Similarly, \citet{Santoro2020} found blueshifted outflow velocities typically from $\sim$500\,kms to $\sim$1000\kms. Although not radio AGN, but similar values have also been found in ionised gas for optically selected type 2 AGN by \citet{Bae2016}, covering the same $L_\textrm{\OIII}$ range as our sample. Our lower outflow velocity for peaked sources tells us that on average, the outflowing gas in radio AGN is not as fast as in the most powerful sources. However, the broad width of the wing component, FWHM = $1330\pm418$\,\kms, also suggests that the peaked sources either have a large variety in outflow velocities, or very turbulent outflows. \par
Our results support the picture where in the early stage of evolution, the radio jets interact with the surrounding medium, piling up energy until eventually they break out, pushing the gas to outflow velocities. As these jets grow, they likely clear a channel of gas in the ambient medium. At the later stages of evolution, there is not a lot of gas in the path of the jet to interact with in the central few kpc region, and push to high velocities, and therefore the outflows detected are not the strongest. In the last few years, hydro dynamic simulations have also suggested a similar scenario, where feedback is most prominent when the jets are newly born and they are strongly interacting with the ambient medium, and weakens as the jets grow beyond the medium (for example, \citealt{Mukherjee2016,Mukherjee2018,Meenakshi2022a}).\par

\subsection{Relevance of other properties}
\label{other properties discussion}

Our mean stacking analysis (see section~\ref{radio classification result} and Table~\ref{model fit}) shows no dependence of the width of the stacked \OIII~profile on radio morphology and luminosity. This could be due to radio morphology and luminosity not being the dominant factor in deciding the impact of the AGN on the ionised gas. This also tells us that these properties do not affect our result on the link with the radio spectral shape. As mentioned in the introduction, previous studies have found that compact sources have the most disturbed gas kinematics suggesting a link between the radio morphology and ionised gas kinematics in AGN. Since the sources are expected to be compact in their early stages of evolution, this provides a link to the AGN life cycle. However, we find that although a large fraction ($\sim$68\%) of the peaked ("young") sources in our sample are compact, a significant fraction ($\sim$32\%) of them are also extended. In some cases, extended sources with a young core could represent multiple epochs of activity. Therefore, we find little evidence for a relation between sources being young and morphologically compact at low frequencies, in our sample. It is worth noting that the low frequency LoTSS maps with LOFAR that we use, are highly sensitive to diffuse emission around these sources, and are therefore well suited for characterising source morphology into compact and extended. \par
The outflow velocity for peaked sources in our sample is lower than what has been found for more powerful radio AGN, such as the the 3C and 2Jy sources (see section~\ref{linktolife-cycle}). This would suggest a relation between the outflow velocity and radio luminosity. However, we do not detect this relation when splitting the group of peaked sources in two groups based on the 1.4\,GHz luminosity. The low number of sources in each group also prevent from obtaining a conclusive result. This will require expanding the sample to higher luminosities, which will allow us to investigate this relationship in more detail. We plan to conduct such a study in the future.\par
The role of radiation pressure in driving feedback in our sample is not entirely clear. In section~\ref{optical classification}, we find that higher ionisation state sources have an outflow component (FWHM = $1208\pm143$\,\kms) in their stacked \OIII~profile. The high ionisation state group also has a larger average $L_\textrm{\OIII}$ by a factor of $\sim$10 (see Table~\ref{sample_properties}). This suggests that radiation pressure also drives the outflow in the optically strongest radio AGN in our sample. However, the lack of difference in the widths of the stacked profiles in \OIII~luminosity groups contradicts this role of radiation. But we cannot conclusively exclude some role of radiation pressure in our sample. In order to understand the significance of radiation pressure in depth, we need to expand the sample to a higher \OIII~luminosity, Eddington ratio, for example. We plan to explore this in a future study.\par
Despite the presence of a broad outflow component in the high ionisation stacked spectrum, we do not expect our result on the link between feedback and radio spectral shape to be driven by radiation pressure. This is because majority of the peaked sources ($\sim$60\%) lie in the low ionisation region on the BPT diagram (see Fig.~\ref{bpt}). Furthermore, a larger fraction of non-peaked sources ($\sim$41\%) lie in the high ionisation region than peaked sources ($\sim$28\%). Peaked sources also have a lower average \OIII~luminosity than non-peaked sources, by 0.3 dex (see Table~\ref{sample_properties}). Therefore, the link between \OIII~profile and the radio spectral shape is due to the properties of the radio jet, and not dominated by radiation pressure.\par

In our analysis of stacked spectra in groups of Eddington ratio (section~\ref{eddingtion ratio classification}) we found no significant difference in the widths of the stacked profiles. This tells us that, on average, the feedback on \OIII~gas in our sample is not driven by the radiation pressure due to accretion. This is contrary to some results found before. For example, \citet{Mullaney2013} used a stacking analysis and found a positive link between the FWHM of the \OIII~profile and Eddington ratio for a sample of optically selected AGN. But it is worth pointing out that their sample, comprised of type 1 AGN,
typically has larger values of $\lambdaup_\textrm{Edd}$ (10$^{-2.5}$-10$^{0}$) and $L_\textrm{\OIII}\,(10^{40}-10^{44}$erg s$^{-1}$). Their estimation of Eddington ratio is also different from our approach. They used the width of the \halpha~line to estimate $L_\textrm{Edd}$ and its luminosity to estimate $L_\textrm{rad}$. More recently, \cite{Singha2021} found that presence of outflows increases with Eddington ratio for a sample of low excitation radio galaxies (LERGs). Although LERGs are comparable to the sources in our sample, they also used the parent sample of \citet{Mullaney2013}, and therefore the same method for estimating the Eddington ratio. Lastly, we point out that the Eddington ratios of our sources also does not affect our result on feedback and radio spectral shape, since the distribution of this ratio is similar for peaked and non-peaked sources, as shown in Fig.~\ref{eddratio distribution}. \par
Finally, studying all the above mentioned properties helps us conclude that the link between the radio spectral shape and feedback on the gas is a result of varying impact of the radio jets on the surrounding medium during their life cycle, and not driven by other radio and optical properties. The lack of an outflow in the stacked profile of non-peaked sources also suggests that extreme feedback, such as an outflow, could also be short lived ($\lesssim$1\,Myr). 
\subsection{Outflow energetics}
\label{outflow energetics}
To asses the effect of the outflow on their host galaxies in peaked sources, we need to estimate its mass and kinetic power. Since in the stacked spectrum, we only detect a blueshifted outflow, the outflow morphology is likely bi-conical on average, where the flux from the redshifted cone is attenuated by the dust in the central few kpc region. For a bi-conical outflow model, assuming constant density clouds with an electron temperature $T=10^{4}K$ and case B recombination, the mass outflow rate ($\dot{M}_\textrm{out}$) can be estimated using the following equation from \citet{Husemann2016}, 
\begin{ceqn}
\begin{align}
    \frac{\dot{M}_\mathrm{out}}{3\,M_{\odot} \mathrm{yr}^{-1}} = \Biggl(\frac{100~ \mathrm{cm}^{-3}}{n_\mathrm{e}}\Biggr)\,\Biggl(\frac{L_{\hbeta}}{10^{41}~\textrm{erg s}^{-1}}\Biggr)\,\Biggl(\frac{v_\mathrm{out}}{100~\textrm{km s}^{-1}}\Biggr)\,\Biggl(\frac{\mathrm{kpc}}{D}\Biggr),
\end{align}
\end{ceqn}
where $n_\textrm{e}$ is the electron density of the gas, $L_{\hbeta}$ is the H$\beta$ luminosity of the wing component, $v_\textrm{out}$ is the velocity of the outflowing gas, and $D$ is the size of the outflowing region. Since we cannot spatially resolve the outflow region, we assume the radial size to be equal to half of the SDSS fibre size, which is 1.5\arcsec. At the average redshift of peaked sources ($z=0.1$), this corresponds to $D=2.8$\,kpc. $L_{\hbeta}$ for the wing component is hard to determine since the signal-to-noise of the H$\beta$ line is much lower than \OIII. Therefore, similar to \cite{Singha2021}, we assume that the wing components of the H$\beta$ and \OIII~line have the same flux ratio as the core components, and estimate $L_{\hbeta}$ by scaling the $L_\textrm{\OIII}$ of the wing component by the flux ratio of the core component. We estimate an $L_{\hbeta,\textrm{wing}}$ of 7.6$\times10^{39}$\,erg\,s$^{-1}$ for the wing component.  \par

For the electron density $n_\textrm{e}$ of the gas, we used the emission line ratio of the \Sii$\mathrm{\lambdaup\lambdaup}$6716,6731\,\AA~doublet. Ideally, we should have only used the emission line ratios of the wing components of the \Sii~doublet, because it would be representative of the outflowing gas. However, it is again extremely hard to do so due to the low signal-to-noise and blending of the doublet. In our stacked spectra for the peaked sources, we do not detect a wing component in the \Sii~doublet. We therefore used the core component to estimate an electron density of $237\pm52\,\textrm{cm}^{-3}$, as discussed in Section~\ref{radiospectralshaperesult}. \footnote{The actual densities of the outflowing gas, estimated using auroral and transauroral lines, could be much higher than estimated using the \Sii$\mathrm{\lambdaup\lambdaup}$6716,6731\,\AA~ratio \citep{Holt2011, Santoro2018, Santoro2020, Rose2018, Davies2020}, leading to significantly lower mass outflow rates.} \par

We estimate the values of $v_\textrm{out}$ using two approaches. The first approach is to assume that velocity of the outflowing gas is the centroid of the wing component, and the broadening is due to the velocity dispersion in the outflowing gas. So we set $v_\textrm{out} = v_\textrm{wing}-v_\textrm{core}$ and $v_\textrm{turb} = FWHM_\textrm{wing}$(deconvolved). This is the most commonly used approach, but it does not account for the line-of-sight projection effects on the observed profile. Therefore, we also used a second approach from \citet{Rose2018}, which assumes that the broadening of the wing component is entirely due to the different projections of the outflowing gas velocities and not the intrinsic velocity dispersion. In this approach, we set $v_\textrm{out} = v_\textrm{max}$, where $v_\textrm{max}$ is the maximum velocity that the outflowing gas can reach, and  $v_\textrm{turb} = 0$. Following \citet{Rose2018}, we calculated the maximum velocity as the velocity at which the cumulative flux of the \OIII~wing component (integrated from low to high velocities) is 5\% of the total flux in wing component. We calculate $v_\textrm{max}=1184$\,\kms from the stacked spectrum.\par

Using the first approach, we estimate a mass outflow rate of $0.09\pm0.04$\,M$_{\odot}$\,yr$^{-1}$, and with the second approach, we estimate a rate of $0.41\pm0.13$\,M$_{\odot}$\,yr$^{-1}$. We would like to note that given the uncertainties in the assumption of the outflow size and density, the true uncertainties in the mass outflow rate are likely to be larger than what we estimate. These mass outflow rates are comparable to what has been found for studies of other radio AGN. \citet{Santoro2020} studied the \OIII~outflows for a sample of nine compact radio AGN, and found outflow rates varying from 0.4 to 20\,M$_{\odot}$\,yr$^{-1}$. For a powerful GPS radio source PKS\,1345+12, \citet{Holt2011} found a total mass outflow rate of 7.7\,M$_{\odot}$\,yr$^{-1}$. Recently, \citet{Speranza2021} studied spatially resolved \OIII~outflows for a sample of 3C radio AGN, and found outflow rates varying from $\sim$1-30\,M$_{\odot}$\,yr$^{-1}$. It is important to remember that all of these studies involved significantly more powerful radio AGN, and studied individual detections of outflows, which is only possible for the more massive outflows. Our stacking analysis probes also weaker outflows that are not detected individually.\par 
Next, we estimate the kinetic power using the method from \citet{Holt2006}, which includes both radial and turbulent components of the outflowing gas. Similar to their approach, we assume that the large width of the outflow component reflects the turbulent motion in the gas. We then used the formula

\begin{ceqn}
\begin{align}
    \dot{E}_\mathrm{kin} = \frac{1}{2}\dot{M}_\mathrm{out}\,\Biggl(v_\mathrm{out}^{2}+\frac{v_\mathrm{turb}^{2}}{5.5}\Biggr),
\end{align}
\end{ceqn}
where $v_\textrm{turb}$ is the deconvolved FWHM of the outflow component. The AGN feedback efficiency ($F$) can then be estimated as
\begin{ceqn}
\label{efficiency}
\begin{align}
      F = \frac{\dot{E}_\textrm{kin}}{L_\textrm{mech}},
\end{align}
\end{ceqn}
where $L_\textrm{mech}$ is the mechanical luminosity of the AGN. We have estimated the mechanical luminosity using the relation of \citet{Cavagnolo2010}, $L_\mathrm{mech}=7.3\times10^{36}(L_\mathrm{1.4\,GHz}/10^{24}\,\mathrm{W\,Hz}^{-1})^{0.7}$ W. This relation was determined by comparing the radio power and the cavity power, for evolved radio jets in giant ellipticals and clusters, that made cavities in the surrounding hot, gaseous haloes as they propagated. \citet{Cavagnolo2010} estimated a scatter of $\approx$0.7\,dex around this relation. To estimate the efficiency required by the radiative luminosity, we replaced $L_\textrm{mech}$ in equation 5 with $L_\textrm{rad}$.\par
We obtain an outflow kinetic power of $1.1\pm0.4\times10^{40}$\,erg\,s$^{-1}$ with the first and $1.8\pm0.6\times10^{41}$\,erg\,s$^{-1}$ with the second approach. This gives a jet feedback efficiency of 0.01\% with the first and 
0.13\% with the second approach. Using the radiative luminosity in equation 5, we obtain similar values for feedback efficiencies.\par%

\begin{figure}
\includegraphics[width=\columnwidth]{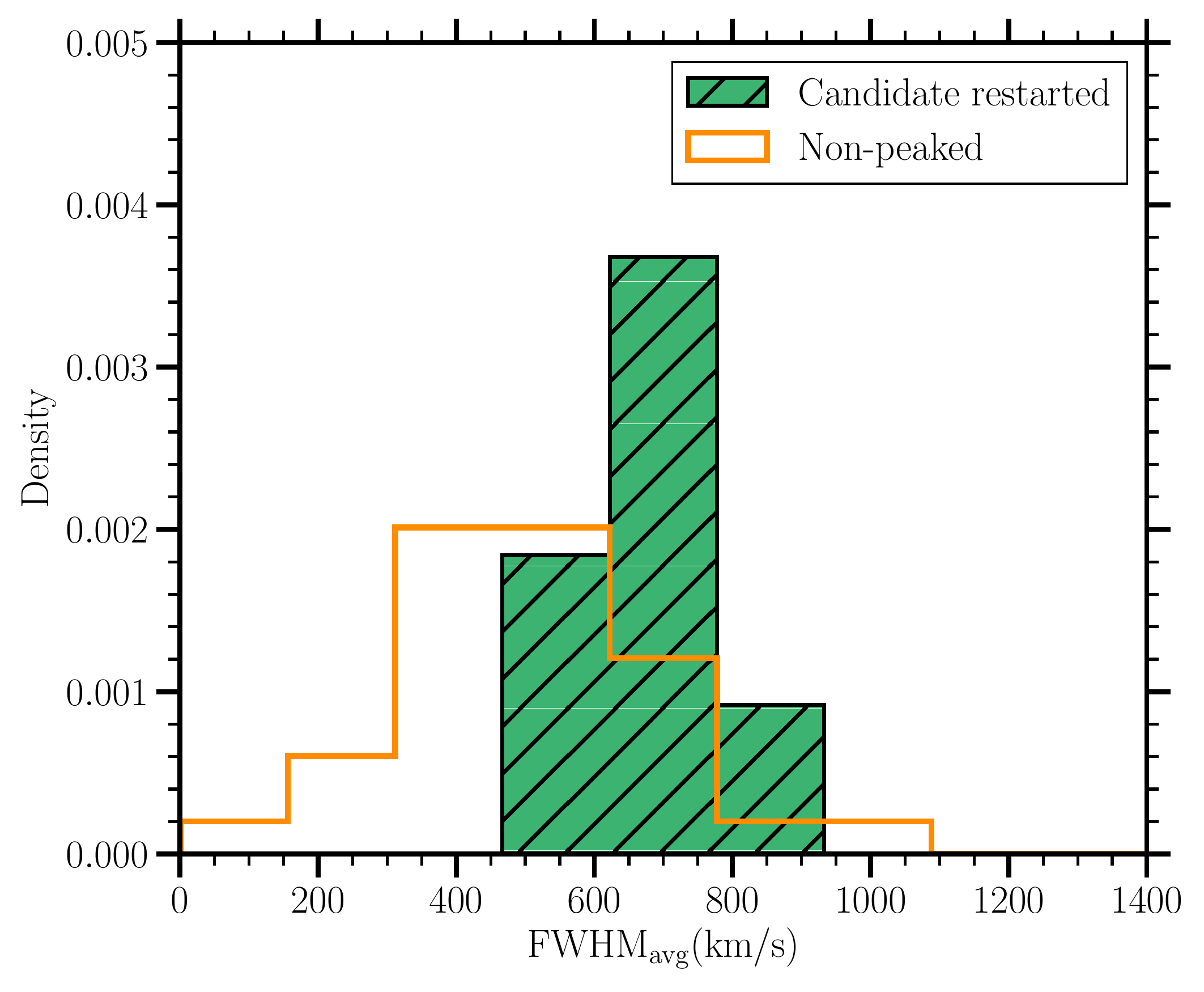}
\caption{ Average FWHM distributions of the \OIII~profiles of candidate restarted and the non-peaked radio AGN.  }
\label{restarted_distribution}
\end{figure}
The feedback efficiencies for our sample of peaked radio AGN are low. Other studies of warm ionised gas in powerful radio AGN have found similar values of feedback efficiencies. For example, \cite{Santoro2020} estimate efficiencies of 0.001\%-0.01\% for the 2Jy sample and \citet{Speranza2021} estimate efficiency of $\sim$0.2\% for some 3C sources. These feedback efficiencies are in agreement with AGN feedback models that estimate $\sim$0.5\% assuming 10\% coupling efficiency \citep{Hopkins2010}, but are significantly lower than the 5-10\% estimated by models that assume 100\% coupling efficiency (for example \citealt{Silk1998,DiMatteo2005}). It is important to remember that our estimates are only for the warm ionised gas phase of the outflow. Majority of the outflowing mass is likely to be in the molecular gas phase, where the outflowing mass can be 2-3 orders of magnitudes higher (for example \citealt{Combes2013,Garcia-Burillo2014,Morganti2015,Fiore2017,Oosterloo2019,Murthy2022}). Therefore, it is crucial to study the multiphase gas to accurately characterise these outflows.

\subsection{Link to episodic activity}
\label{restarted section}

As mentioned in Section~\ref{radio morphology}, we find eight sources that show extended emission only in their LoTSS images (144\,MHz) and not at 1.4 and 3\,GHz. The lack of extended emission at higher frequencies indicates that the radio spectral index of this emission is steep, possibly due to synchrotron losses. This suggests that the extended emission is old. Out of these sources, seven have a peaked spectrum from 144-3000\,MHz, suggesting the presence of a young radio AGN in the central region. We used the combination of these two properties to classify these seven sources as candidate restarted radio AGN. \par
Out of the seven candidates, six have \OIII~detections, and we plot the distribution of their FWHM$_\textrm{avg}$ in Fig.~\ref{restarted_distribution}, along with the non-peaked sources. These candidates seem to have a broader \OIII~profile than the non-peaked sources. A two sample KS test gives a statistic of 0.59 and a p value of 0.03. The low p value suggests that the two distributions are significantly different. This tells us that restarted sources could have more disturbed gas kinematics than evolved radio AGN. However, we note that the number of candidates in our sample are too low to conclude this with confidence. Finding concrete evidence for this trend would require detailed radio spectral studies to confirm the restarted nature of activity in the candidates. Nevertheless, this trend is in line with what we expect from our results, since restarted sources have a young radio AGN in the centre. In the picture we describe above, this would mean that after the jets switch off, the gas falls back in to the regions that were cleared, and therefore when the jets restart, there is gas again available for interaction. This suggests a link between feedback on \OIII~gas and episodic activity of radio AGN.
\section{Conclusions and future prospects}
\label{summary}
Using a combination of multi-frequency radio data to constrain the radio spectral shape, and optical spectra to model the ionised gas kinematics and AGN properties, we have linked the life cycle for a sample of 129 radio AGN to their impact on the surrounding ionised gas. The stacking analysis allowed us to disentangle the role of different AGN properties. Our main results are summarised below:
\begin{enumerate}
    \item We find that over a radio luminosity range of $\sim$10$^{23}$-10$^{26}$\whz{} and up to $z\sim0.2$, the radio spectral shape is linked to the feedback on the ionised gas. Peaked sources have the most disturbed ionised gas kinematics, with a turbulent outflow of $v_\textrm{out}$\,$\approx$\,$-237\pm102$\,\kms and FWHM = $1330\pm418$\,\kms in their stacked spectrum. 
    
    \item Feedback on the ionised gas evolves over the radio AGN life cycle, and radio jets are most effective at disturbing the gas in the ambient medium when they are young. Lack of an outflow in the stacked spectrum of non-peaked sources suggests that these outflows are short lived.

    \item The difference in the stacked profiles of peaked and non-peaked sources is not driven by radio properties such as the morphology and luminosity, and optical properties such as the ionisation ratio, radiative luminosity, and Eddington ratio. 
    
    \item On average, the mass outflow rate is $0.09-0.41$\,M$_{\odot}$\,yr$^{-1}$ and kinetic power is $0.1-1.8\times10^{41}$\,erg\,s$^{-1}$ for outflows in peaked sources. These values are small, but comparable to those found for other more powerful radio AGN. 
    
    \item Although not conclusive, the presence of broader \OIII~profiles in candidate restarted AGN in our sample than the non-peaked sources suggests a link between feedback and the episodic activity of radio AGN.
\end{enumerate}
It is worth investigating the impact of this kind of AGN driven feedback on the host galaxy, which, as we mention above, is likely short lived. The small mass outflow rates also tell us that these outflows are likely not an efficient way to quench star formation in the host galaxy. However, jet-ISM interaction is also expected to make the ambient medium turbulent. This turbulence could keep the gas clouds from collapsing to form stars. Therefore, it would be interesting to explore a larger sample of radio AGN and their optical spectra to look for signatures of jet induced turbulence, and how that evolves with the life cycle. This could be the long term feedback on the gas, and might be more efficient at quenching star formation. We would also detect more candidate restarted radio AGN, thus investigating the link of feedback with episodic activity in more detail. We are currently working on a project to expand this sample to higher redshifts and luminosities. Although we use unresolved SDSS spectra, future studies with integral field unit observations would be crucial to study this form of feedback in detail, and evaluate its importance in the galaxy's evolution.

\begin{acknowledgements}
We would like to thank the anonymous referee for the comments, that helped improve this paper. LOFAR data products were provided by the LOFAR Surveys Key Science project (LSKSP; https://lofar-surveys.org/) and were derived from observations with the International LOFAR Telescope (ILT). LOFAR \citep{vanHaarlem2013} is the Low Frequency Array designed and constructed by ASTRON. It has observing, data processing, and data storage facilities in several countries, which are owned by various parties (each with their own funding sources), and which are collectively operated by the ILT foundation under a joint scientific policy. The efforts of the LSKSP have benefited from funding from the European Research Council, NOVA, NWO, CNRS-INSU, the SURF Co-operative, the UK Science and Technology Funding Council and the Jülich Supercomputing Centre. The National Radio Astronomy Observatory is a facility of the National Science Foundation operated under cooperative agreement by Associated Universities, Inc. CIRADA is funded by a grant from the Canada Foundation for Innovation 2017 Innovation Fund (Project 35999), as well as by the Provinces of Ontario, British Columbia, Alberta, Manitoba and Quebec.

\end{acknowledgements}
  \bibliographystyle{aa-copy} 
  \bibliography{References} 
\onecolumn

\begin{appendix}
\section{Continuum images}
\begin{figure}[!ht]
  \includegraphics[width=\columnwidth]{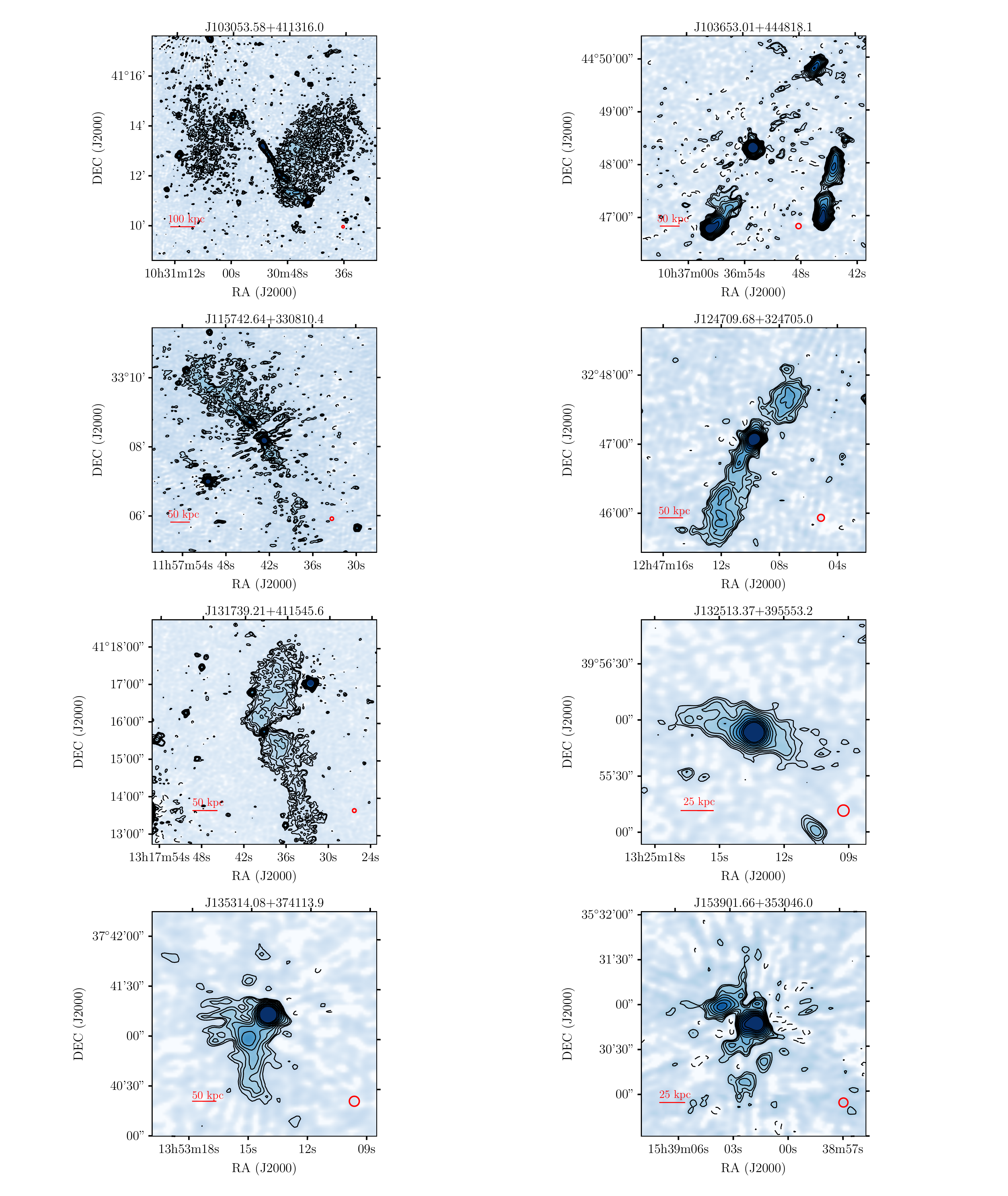}
  \caption{\small LoTSS continuum images of eight sources that were classified are compact in their FIRST images, but show clear signs of extended emission at 144\,MHz. The contour levels in all images are 3$\sigma_\mathrm{RMS}\times\sqrt{2}^{n}$ where n=0,1,2... and $\sigma_\mathrm{RMS}$ is the local RMS noise in the images. The negative contours are marked by dashed lines and are at $-3\sigma_\mathrm{RMS}$ level. All of these sources (except J103653.01+444818.1) have a peaked source in the centre and are classified as candidate restarted sources. }
  \label{new_extended}    
\end{figure}

\section{Model fits for stacked spectra}
\begin{figure}[!ht]
  \includegraphics[width=\columnwidth]{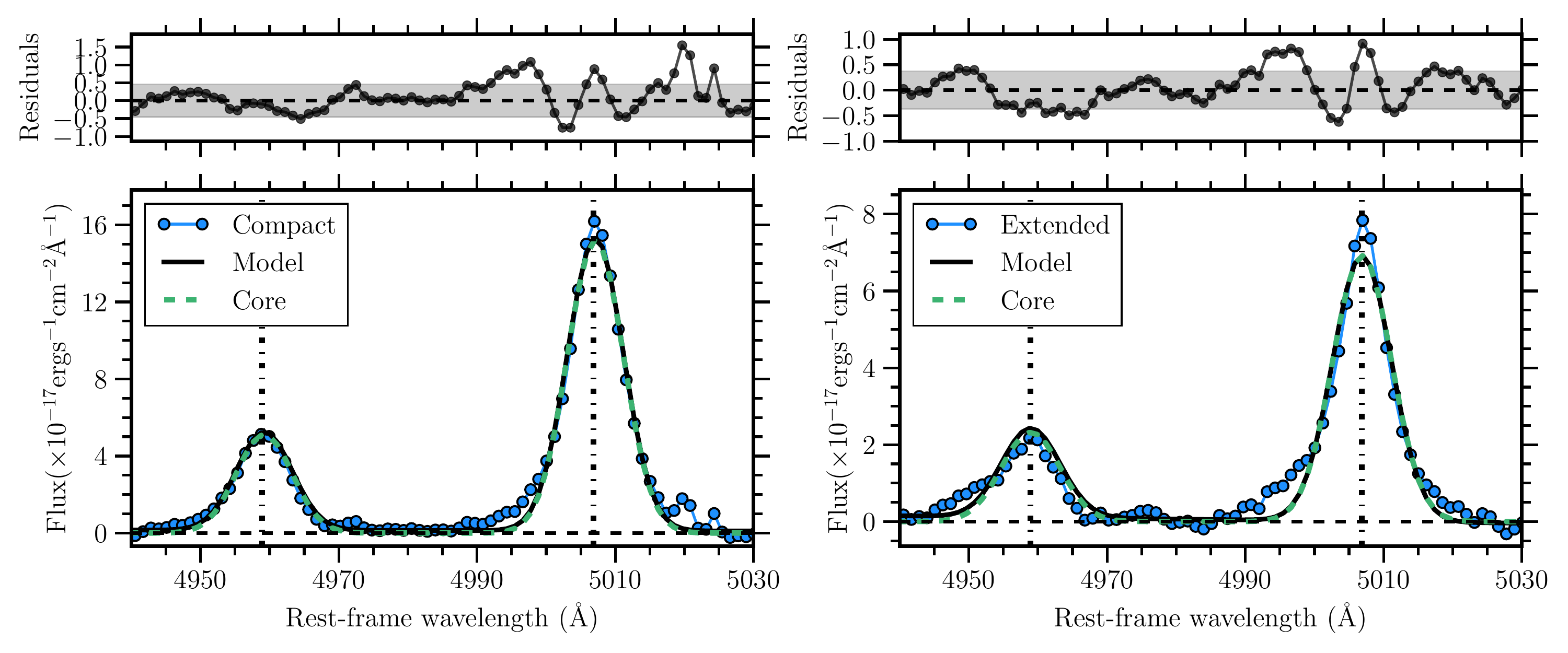}
  \includegraphics[width=\columnwidth]{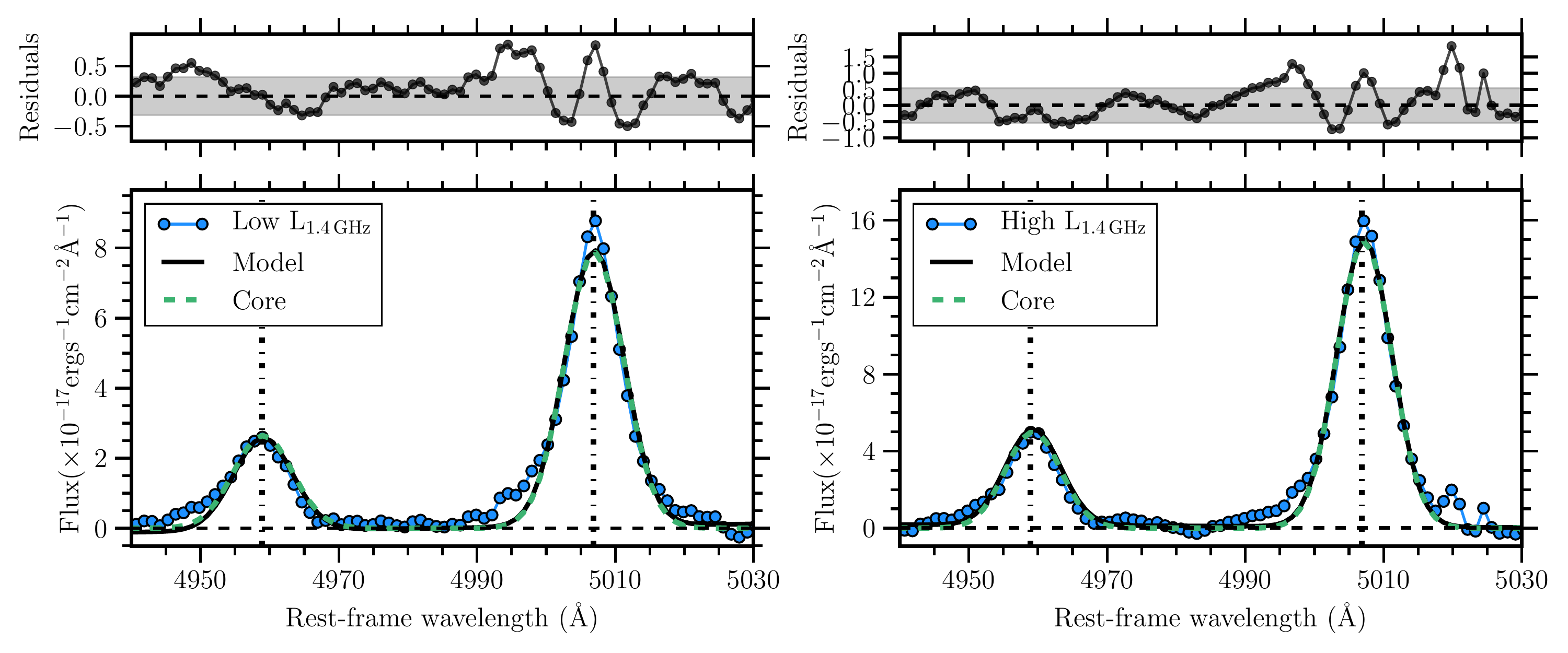}
  \includegraphics[width=\columnwidth]{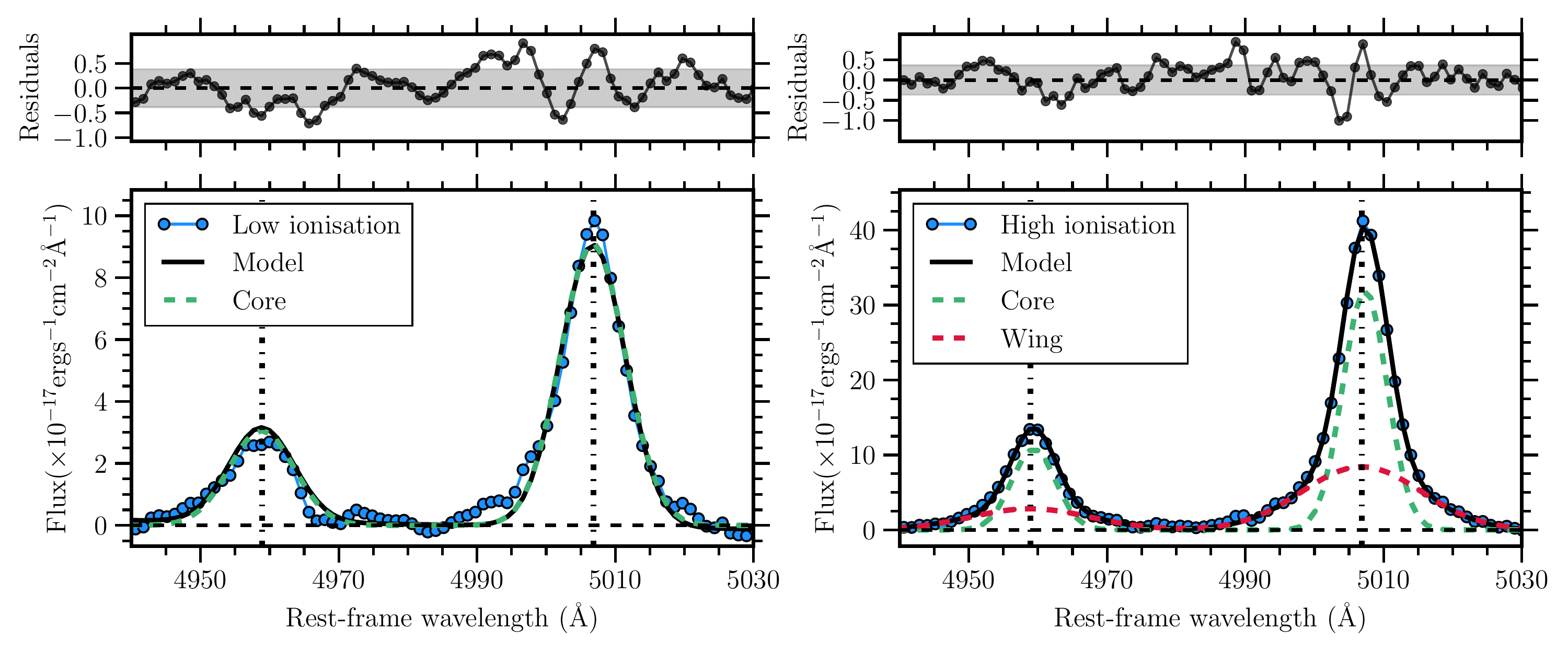}
\end{figure}
\begin{figure}[!ht]
  \includegraphics[width=\columnwidth]{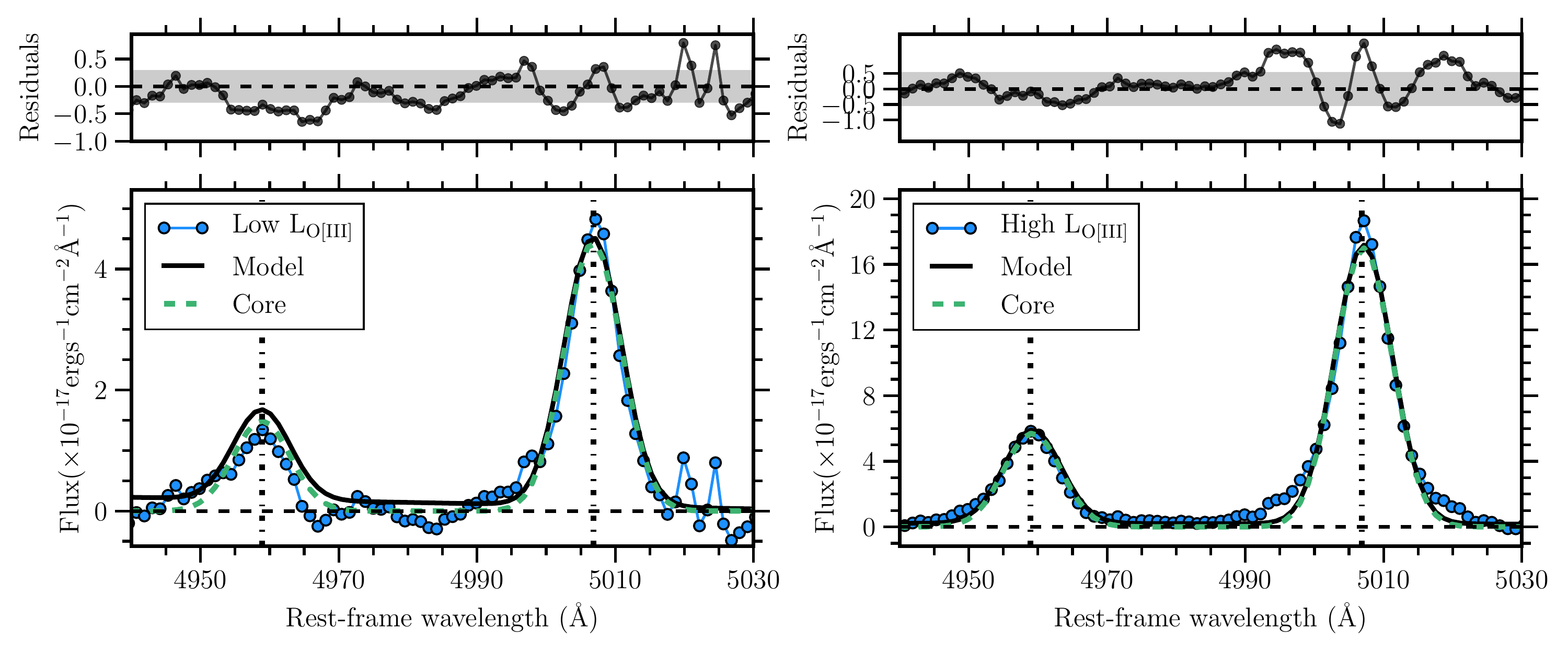}
  \includegraphics[width=\columnwidth]{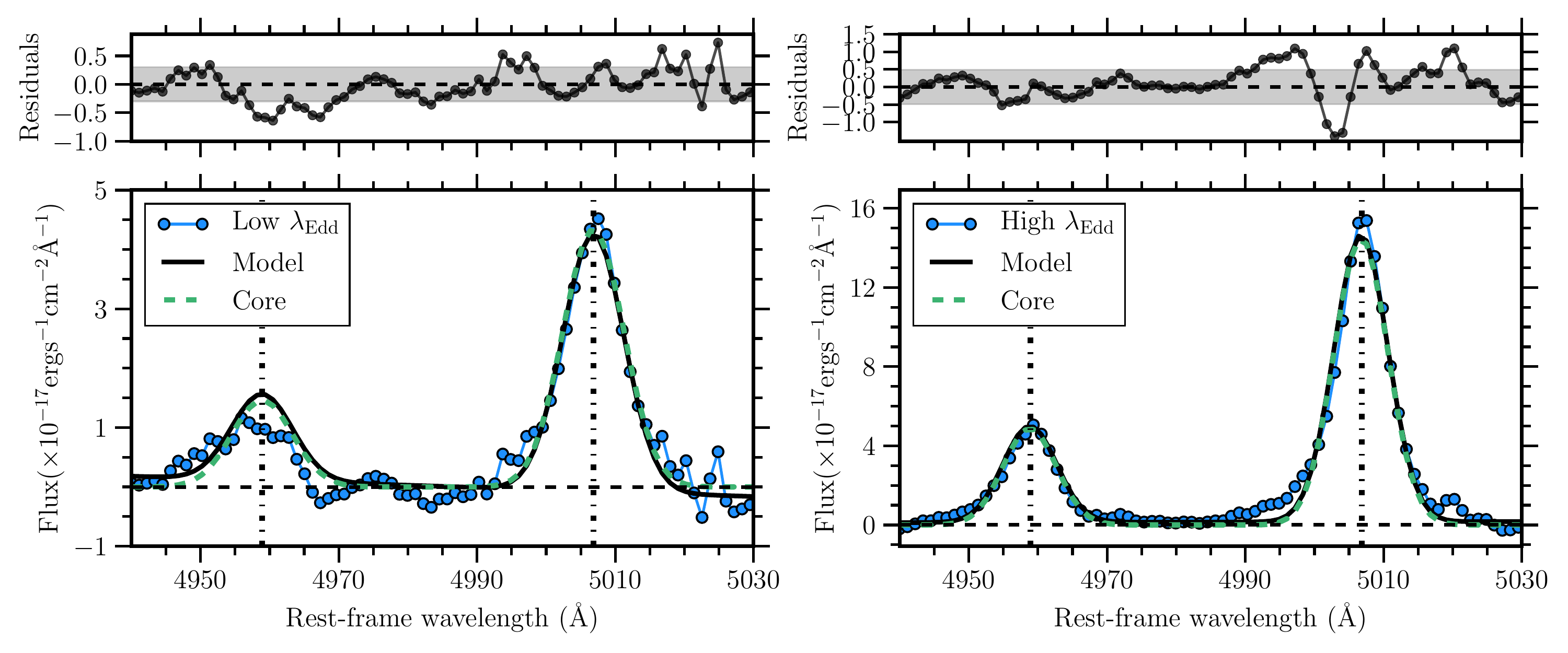}
  \begin{center}
  \includegraphics[width=0.5\columnwidth]{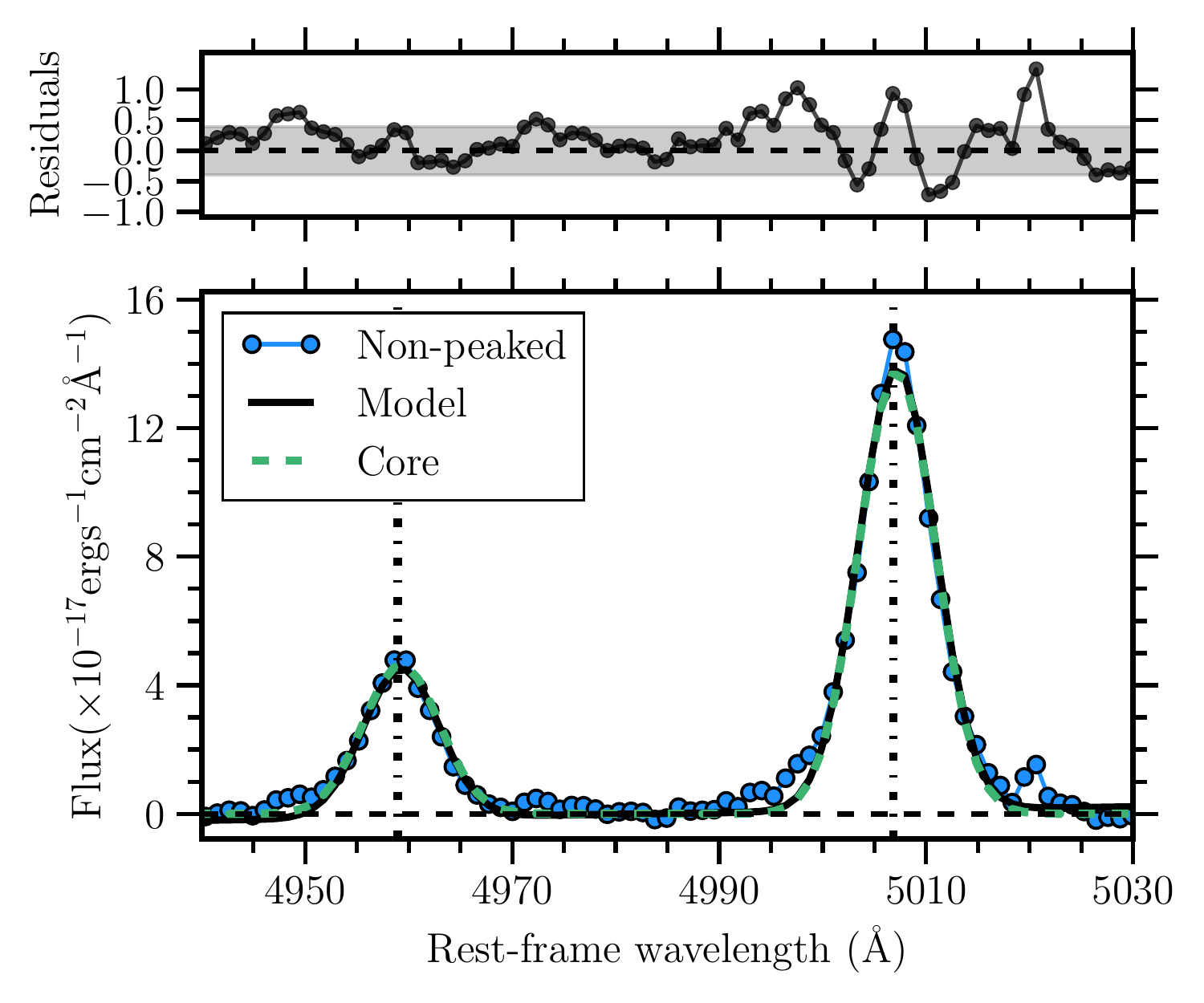}
  \end{center}
  \caption{\small Model fits to the stacked \OIII~spectra of various groups. Bottom panel shows the data, the best fit model and the individual components. Top panel shows the residuals after subtracting the best fit model from the data and. Model parameters are given in Table~\ref{model fit}. }
  \label{modelfits_spectra}    
\end{figure}
\begin{figure}[!ht]
  \includegraphics[width=\columnwidth]{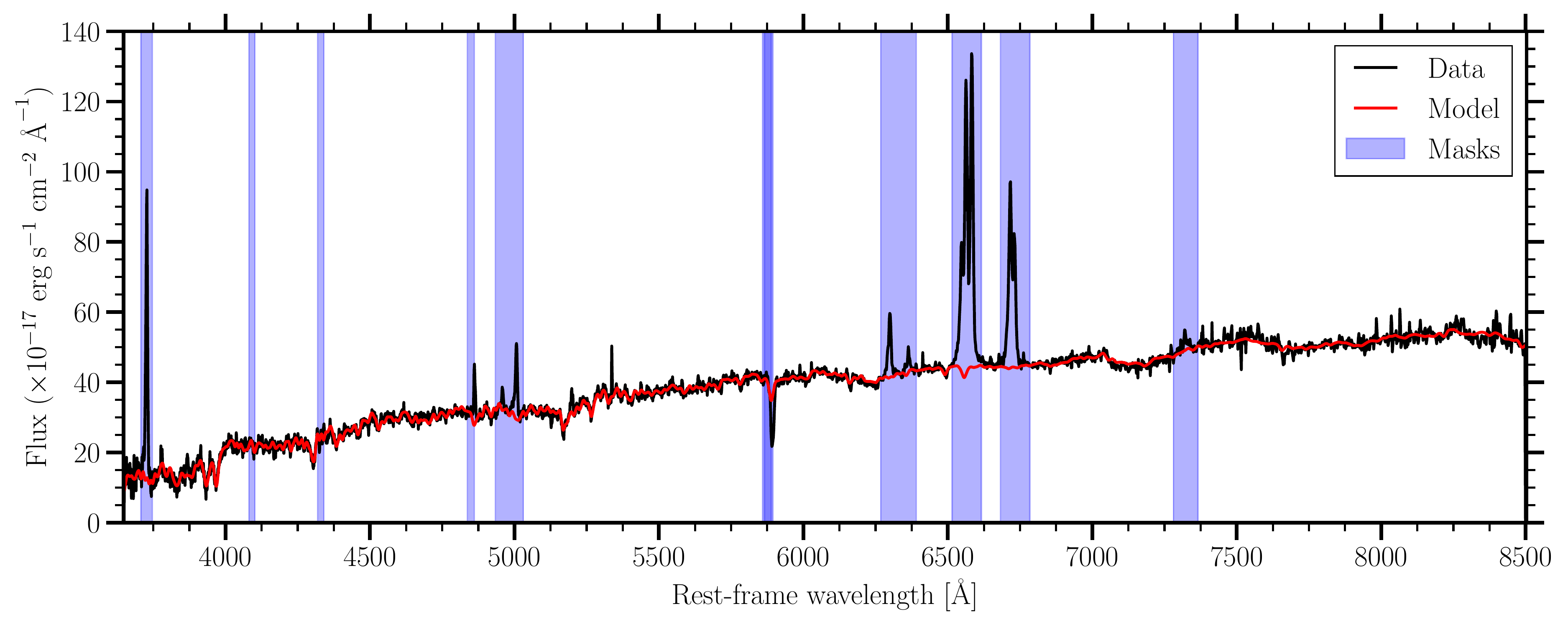}
  \caption{\small An example of the SDSS spectra and stellar continuum fit for the source J135217.88+312646.4. The blue regions show the emission lines masked during the fitting. The model reproduces the stellar continuum of our data very well.}
  \label{stellarcontinuum}
\end{figure}
\end{appendix}
\end{document}